*Article – original research*

# Nap-induced modulations of tinnitus – a cross-sectional database analysis

**Robin Guillard [1,\*], Martin Schecklmann [2,3], Jorge Simoes [2,3,4], Berthold Langguth [2,3] Alain Londero [5], Marco Congedo [1], Sarah Michiels [6], Markku Vesala [7], Hazel Goedhart [7], Thomas Wetter [8], and Franziska C. Weber [2,3\*]**

[1] GIPSA-Lab , Univ. Grenoble Alpes, CNRS, Grenoble INP, Grenoble, France.
[2] Department of Psychiatry and Psychotherapy, University of Regensburg, 93053 Regensburg, Germany
[3] Interdisciplinary Tinnitus Centre, University of Regensburg, 93053 Regensburg, Germany
[4] Faculty of Behavioural, Management and Social Sciences, University of Twente
[5] APHP, Hôpital Européen Georges-Pompidou, Service ORL et Chirurgie Cervico-Faciale, APHP Paris, France;
[6] Department of Rehabilitation Sciences and Physiotherapy, Faculty of Medicine and Health Sciences, University of Antwerp, Wilrijk, Belgium
[7] Tinnitus Hub Ltd, Hemsworth, United Kingdom

\* Correspondence: robin.guillard@grenoble-inp.fr; Franziska.Weber@medbo.de

**Abstract:** It is clinically well known that in some individuals tinnitus is positively or negatively influenced by naps. Here we aimed at systematically investigating this phenomenon, by assessing its frequency and by exploring associated clinical or demographic characteristics.

Data from two different databases with a total of 9,742 people with tinnitus were investigated. Through the Tinnitus Hub "Investigating physical links to tinnitus" survey database, 6,115 participants with tinnitus were sampled for a comprehensive survey regarding tinnitus characteristics. Using the TRI database, 3,627 outpatients who presented to a German tertiary tinnitus clinic during the period between 2010 and 2023 and had completed validated tinnitus questionnaires were included. After separate analysis of the databases, these results were then compared with each other.

In the Tinnitus Hub survey database, a total of 31.1% reported an influence on tinnitus by taking a nap, with much more frequent worsening after a nap than improvement (23.0% a little or a lot worse, 8.1% a little or a lot better). In the TRI database, a total of 26.9% reported an influence on tinnitus by taking a nap (17.7% worse, 9.2% better).

The influence of napping on tinnitus was associated in both databases with other clinical features, such as the dependence of tinnitus on night quality, stress and somatosensory maneuvers.

The present study confirms the clinical observation that more tinnitus sufferers report worsening after a nap than tinnitus sufferers reporting an improvement. It was consistently shown that tinnitus sufferers reporting nap-induced modulation of tinnitus also report more frequently an influence of night sleep on their tinnitus. Co-occurence of influence of stress on tinnitus and influence of naps on tinnitus could suggest a specific influence of serotonergic drive on central auditory pathways. Whether the observed association between tinnitus worsening after naps and somatosensory modulation of tinnitus might be mediated by the occurrence of sleep associated symptoms such as bruxism, snoring or sleep apnea events during naps should be investigated in future studies. Further clinical and polysomnographic research is warranted to better understand the interaction between sleep and tinnitus.

## 1. Introduction

Tinnitus is defined as the perception of a sound in the absence of an external acoustic source [1,2]. The prevalence is high at about 10-15%, with about 1-2% of the general population also having severe tinnitus [3–8]. Tinnitus shows a high clinical heterogeneity, e.g., regarding loudness, pitch, sound quality, laterality and continuity [2]. While some tinnitus patients have constant tinnitus, some have variable tinnitus, e.g., intermittent or fluctuating in intensity [9]. Tinnitus patients often report that some factors can modulate their tinnitus, causing either an improvement or a worsening.

In this context, the clinical phenomenon that a daytime nap can have an effect on tinnitus is well known. In clinical settings, this phenomenon is queried in the frequently used Tinnitus Sample Case History Questionnaire (TSCHQ) of the Tinnitus Research Initiative (TRI) with question 22 (influence by nap: worsens/improves/no effect) [10], as well as in the ESIT-SQ questionnaire of the European School of Interdisciplinary Tinnitus research (ESIT) as a potential modulatory factor of tinnitus (question B18 on factors decreasing tinnitus and question B19 on factors increasing tinnitus) [11]. It is remarkable that individually different effects are possible and also that a measure intuitively assessed as relaxing can have negative effects on tinnitus. To date, this phenomenon has only been rudimentarily investigated scientifically [12–14] and sometimes compared to the rare parasomnia figuratively called "exploding head syndrome" due to the fact in both cases, patients describe abnormally loud sounds upon awakening [15,16]. Several past reviews suggested that nap modulations of tinnitus were a specific subcase of somatosensory modulation of tinnitus and reported anecdotally that this phenomenon seemed to be amplified in the sitting position [15–17]. A recent study investigated night polysomnographic characteristics of tinnitus patients reporting increases of tinnitus after naps. It compared them to the characteristics of a matched control group of tinnitus patients not presenting such increases. Patients with nap-induced tinnitus increase seemed to have less deep sleep (N3) and less paradoxical sleep (Rapid-eye movement / REM) than controls [18].

Yet a symptomatologic characterization of the population of tinnitus patients reporting an effect of naps on their tinnitus remains elusive. To fill this gap, samples from two ecologically relevant large databases (from the TRI database [19] and from the online forum Tinnitus Talk https://www.tinnitustalk.com// [20]) were used to clinically characterize groups of patients who experience an improvement, those who perceive a worsening and those who deny any influence of naps on their tinnitus perception. This approach aims (1) at a better understanding of the influence of sleep on tinnitus and (2) at exploring whether nap-induced modulation might represent a clinically useful criterion for tinnitus subtyping.

## 2. Materials and Methods

### 2.1. Data collection

#### 2.1.1. TRI database

The TRI database consists of a sample of 3627 questionnaire records from patients who visited the tertiary tinnitus clinic between 2010 and 2023 [19]. All data were collected at intake assessments during the first visit at the clinic. This decision was made to mitigate the potential biases of clinical interventions and/or multiple visits to the clinic on the target variable, in the present case, nap-induced tinnitus modulations. Ethical approval for the collection of this database was obtained from the Ethics Committee of the University of Regensburg, protocol number 08/046, and all patients provided their written consent after oral information to have their data pseudonymised, stored, analyzed, and published for scientific purposes.

2.1.2. Tinnitus Hub "Investigating physical links to tinnitus" survey database

The Tinnitus Hub survey database consisted of a sample of 6115 records from participants collected through a Web-based survey in February 2017. The initial goal of this survey was to assess the presence of physical symptoms in a convenience sample of participants with tinnitus. It was launched on the online forum, Tinnitus Talk, managed by Tinnitus Hub.

As presented in [20], the motivation of the survey topic emerged from participants' interactions on the online forum and is a good example of citizen science [21]. Questions were designed in consultation with tinnitus researchers and a small pool of the forum's community and trialed with this group before launch. This was done to make sure that all questions were clear and unambiguous and that no technical issues were present. The final questionnaire consisted of 21 questions that asked for different physical symptoms that come as comorbidities with tinnitus and for a set of tinnitus characteristics. Among these comorbidities, the participants were asked whether or not naps modulated their tinnitus.

The survey was advertised on the Tinnitus Talk website and launched as an open survey, on the online forum, Tinnitus Talk. No incentives were offered to participants. An internet protocol check was used to identify and block potential duplicate entries from the same user. All participants gave informed consent for their anonymized data to be used for scientific purposes. No personal information was collected during the process. Ethical approval was obtained from the Ethical committee of the Antwerp University Hospital for the use of this survey for research purposes (Reference number : 19-43-485).

*2.2. Databases variables*

In the present study we analyzed two databases reaching different subpopulations of tinnitus patients, to reduce selection bias and to obtain more robust results [22] . The questions about nap-induced tinnitus modulation were quite similar in the two databases: "Does taking a nap during the day affect your tinnitus ?" for the TRI database with possible answers "worsens my tinnitus", "reduces my tinnitus" and "has no effect". In the Tinnitus Hub database, the question was "How do the following affect your tinnitus : waking up from napping ?". Possible answers were "A lot worse", "A little worse", "No difference", "A little better", "A lot better". Some other features of these databases were in common and presented hereafter. Also, some features were only present in the TRI or the Tinnitus Hub databases, they will also be presented hereafter.

2.2.1. Common variables to the two databases

Common variables in the two databases in addition to nap modulations were :
- Gender and age of participants;
- Tinnitus duration;
- Whether tinnitus consisted in a pulsatile sound, a tonal sound or another type of sound;
- Events related to tinnitus onset;
- Rating scales of tinnitus loudness and annoyance;
- Percentage of time aware of the tinnitus percept;
- Influence of stress, sleep (different formulations) and noise (possibility to mask it, worsening) on tinnitus;
- Presence of somatosensory modulations of tinnitus
- Self-reported hearing loss grade;

- Reported comorbidities in addition to tinnitus: headache, neck pain/stiffness (different formulations), temporomandibular pain (different formulations).

2.2.2. TRI database specific variables

Variables unique to the the TRI database included :

- Handedness of participants
- Laterality of the tinnitus percept;
- Whether tinnitus appeared progressively or abruptly;
- Whether or not other direct family members had tinnitus;
- Standardized questionnaire scores such as the Tinnitus Handicap Inventory (THI) and the Tinnitus Questionnaire (TQ);
- More granularity on tinnitus grading scales on how much the participant is bothered by tinnitus through the "Tinnitus Severity" questionnaire;
- Impact of tinnitus on the general quality of life measured by the World Health Organization Quality Of Life (WHOQOL-bref) questionnaire on different dimensions;
- Percentage of time participants felt angry and/or frustrated by their symptoms
- Other reported comorbidities in addition to tinnitus such as, e.g., vertigo or, hyperacusis
- How many treatments patients had already tried to alleviate their tinnitus;
- Whether or not patients are using hearing aids and/or receive psychiatric treatment because of their tinnitus

2.2.2. Tinnitus Hub survey database specific variables

Variables unique to the Tinnitus Hub survey database included :

- More granularity on the variability of tinnitus loudness and annoyance over time;
- Whether patients experienced fleeting tinnitus events and on which frequency;
- Events that can trigger ear fullness
- Whether or not patients tend to clench or grind their teeth
- Tinnitus modulations related to jaw movements;
- More granularity on the way the jaw or the temporomandibular joint is aching;
- Influence of anxiety and sport practice on the tinnitus percept

*2.3. Data processing*

Data processing consisted first in labeling all missing answers similarly, notably, when patients responded "I don't know" when this choice was possible, it was counted as a missing answer.

Categorical variables encoded as strings in response to multiple choice questions were treated as separate columns and each of the created columns received a value of "1" if the answer choice was selected, 0 if not.

Another processing step for the TRI database consisted in excluding all records that were not a first visit from a patient (i.e., all follow-up visits records). Moreover, only scores and not individual question answers from standardized questionnaires THI, TQ, Tinnitus Functional Index (TFI), Major Depression Index (MDI), WHOQOL-bref were included as variables. As TFI and MDI scores were only available in less than half of the patients, they were not included in the analysis.

There was a difference in the possible answers to the question of nap-induced modulations of tinnitus between the two databases. In fact, in the TRI database, there were three choices : "worsens my tinnitus", "reduces my tinnitus" and "has no effect" whereas in the Tinnitus Hub survey database, there were 6 possible answers : "A lot worse", "A little worse", "No difference", "A little better", "A lot better" and "Unsure / don't know" (the records who selected this last option were excluded from the analysis).

The difference in granularity of the answers is reflected in the Figure 1 and 2 of the Results section. Yet, in order to be able to perform the same group comparisons on the two databases, in the Tinnitus Hub survey database the answers "A lot worse" and "A little worse" were merged in one column "Worse" and similarly the columns "A lot better" and "A little better" were merged in a column "Better".

As the possible answers for stress, anxiety, good sleep, bad sleep and light, moderate and intense workout influence on tinnitus in the Tinnitus Hub survey database were saved with the same possible answers choices, the same processing (merging "A lot worse" and "A little worse" and merging "A lot better" and "A little better") was applied to these variables as well.

*2.4. Statistical analysis:*

The two databases were analyzed separately. The analysis consisted in comparing in each database the differences between the three groups : "Tinnitus Worsened" by naps (TW), "No Effect" of naps on tinnitus (NE) and "Tinnitus Improved" by naps (TI).

There were four types of answers to questionnaire items : scales (for example, tinnitus annoyance rating scale), gradations (for example for the degree of self reported hearing loss), categorical unique choices (only one choice of answer could be selected among a set of possible answers, as for example for tinnitus laterality) and finally categorical multi-choices (the participant could select as many answers as he wished among a selection of possible answers, as for example the sound characterization of the tinnitus percept). Ratings and gradations were analyzed similarly and referred to in the following sections as scales.

2.4.1. Variable testing

For scale type of answers, the comparison between the three groups was performed with the non-parametric Kruskal Wallis test. An effect size was associated with this Kruskal Wallis test, an implementation of this effect size can be found in the R.statix 0.6.0 package at the following link : https://www.rdocumentation.org/packages/rstatix/versions/0.6.0/topics/kruskal_effsize . For categorical unique choices type of answers, a Chi-Squared test was performed between all possible answers. For categorical multi-choices, each possible answer was assigned as an independent boolean variable and was tested separately between "Yes" and "No" with a Chi-Squared test as well. For all these categorical types of answers, Cramer's V was selected as effect size.

2.4.2. Post-hoc tests

For questions with a scale type answer, Dunn's non-parametric test [23] was used for pairwise post-hoc comparisons. As no specific effect size formula could be found associated with the Dunn test statistic, Cohen's d was calculated as effect size for each pairwise test. For categorical types of answers, pairwise chi-squared tests were performed and Cramér's V was used as effect size for each pairwise test.

2.4.3. Control for multiple comparisons

Holm adjustment for multiple comparisons [24] was applied on all variables before post-hoc testing for each database independently. Pairwise post-hoc tests were only performed if the controlled p-values of the tested variable were below 0.05. A second Holm adjustment for multiple comparisons procedure was then applied on the obtained post-hoc p-values of all post-hoc tests for each database independently.

*2.4. Software choices*

Data processing and statistical analyses was performed in R (version 4.0.3, R Core Team, Vienna, Austria, 2020) and in Python (Anaconda 2022.10, Austin, Texas, USA and PyCharm 2020.3.2, JetBrains, Prague, Czech Republic). Data processing in R was performed with the packages tidyverse [25] and tidymodels with its default settings [26]. Data processing and statistical analyses were performed in python with the Pandas library [27] the scipy.stats library [28], the statsmodel library [29] as well as the Scikit post hoc library [30].

The code used to perform this analysis is accessible on the GitHub open-source repository at this URL : https://github.com/RobinGuillard/TinniNap_DB_study

The TRI data presented in this study are available on reasonable request
 from the corresponding author.
The Tinnitus Hub survey database is available on reasonable request to Tinnitus Hub and S.M.

**3. Results**

*3.1. Nap influence on tinnitus in the two databases*

In the Tinnitus Hub survey database, 31.1% reported an influence on tinnitus by taking a nap (23.0% a little or a lot worse, 8.1% a little or a lot better), in the TRI database 26.9% (17.7% worse, 9.2% better) (Figure 1 and 2). There are no significant differences between the distributions of participants' modulation of tinnitus through naps in the two databases (p = 0.99).

The TW group was found to be much more numerous than the TI group (TW : 1404 VS TI : 507 more than double in the Tinnitus Hub survey database, TW : 642 VS TI : 334 not far from the double in the TRI database).

The main socio-demographic data and tinnitus characteristics with distribution differences of the two databases can be found in Table 1. This comparison shows that on the majority of common criteria in both databases significant differences exist. The most striking difference can be observed for self-reported stress being the initial cause of tinnitus onset : a medium effect size difference of 0.35 was observed between the TRI database (50,6% yes, 49,4% no), and the Tinnitus Hub survey database (17% yes, 83% no). To a lesser extent, small effect size differences between databases samples were identified for age (higher in TRI Database), tinnitus loudness (higher in TRI Database), and annoyance (higher in TRI Database) scales, gender (more males in TRI Database), acoustic trauma as the initial cause of tinnitus onset (higher in Tinnitus-Hub Database), hearing impairment (higher in Tinnitus-Hub Database), somatosensory modulations (higher in Tinnitus-Hub Database) and modulation of tinnitus by sounds worsening (higher in TRI Database) and suppressing (higher in Tinnitus-Hub Database).

*3.2. Group comparisons*

Group comparison in the TRI database and the Tinnitus Hub Survey Database are respectively presented in tables 2 and 3. Only variables eliciting significant differences after Holm correction were reported. For comparison on the complete list of variables, please refer to Supplementary Table 1 and 2 in the Supplementary Material.

3.2.1. Congruent findings

In both databases, the largest effect sizes for group comparison were obtained for the variables concerning sleep. In the TRI database, the relationship between sleep and tinnitus during the day was stronger in TI and TW than in NE, with an overall medium effect size. Likewise, similar results were found in the Tinnitus Hub survey database, where the influence of a good night sleep elicited important differences between the 3 groups, with again the NE group standing out, as well as the influence of poor sleep.

Then, in both databases, modulation by stress and by somatosensory maneuvers seemed to elicit consistent small effect size comparison differences.

Finally to a lesser extent, in both databases, congruent significant differences between groups were found on gender, age, tinnitus annoyance, initial cause of onset linked to a stressful situation and influence of sound on tinnitus : worsening of tinnitus with sounds in some situations, and possibility to mask tinnitus with specific type of sounds.

On the other hand, in both databases as shown in the Supplementary Material, several common features were consistently found not to differentiate between groups : acoustic trauma and presbycusis as initial causes of tinnitus onset, neck and temporomandibular pain prevalences, headache prevalence, tonal or pulsatile forms of tinnitus prevalences.

3.2.2. Differences between databases

The only notable incongruence between database group comparisons was found on tinnitus duration comparison. Group comparison was significant for the Tinnitus Hub survey database with the "No effect of naps" group having the longest tinnitus duration. However, in the TRI database, group comparison was not significant and the group reporting worsening of tinnitus following naps was found to have a longer duration than the "No effect of naps" group.

Additionally, on several variables, only one database yielded significant differences in group comparison. In fact, a significant difference on tinnitus loudness was found only in the TRI database, although the trend tended to point in the same direction in the Tinnitus Hub survey database.

Moreover, a significant difference in the percentage of time aware of the tinnitus percept, head or neck injury as initial causes of onset of tinnitus and hearing loss were only found in the Tinnitus Hub survey database although, again, the trend pointed in the same direction in the TRI database.

3.2.3. Findings in variables unique to TRI database

Among the variables only explored using the TRI database, group comparison was found significant on the side of the tinnitus (with less lateralization for TW compared to TI and NE, above all in the left ear), although results were found not significant for the degree of lateralization. Group comparison also found differences on whether tinnitus was intermittent or continuous (with more intermittent tinnitus in TI and TW compared to NE group), whether it fluctuated from day to day (more variations for TI and TW groups compared to NE group), THI and TQ questionnaire scores (more tinnitus distress in TW group compared to NE group), hyperacusis (more hyperacusis for TI group compared to TW and NE group), physical discomfort when exposed to noise (more discomfort for TI and TW groups compared to NE group), number of treatments tested by participants (more tinnitus treatments tested for TW group compared to NE group), the proportion of participants who tried using hearing aids (more patients of TW group tested hearing aids compared to NE group).

3.2.4. Findings in variables unique to Tinnitus Hub survey database

Among the variables only explored by the Tinnitus Hub survey database, group comparison was found significant on ototoxic treatment as the initial cause of onset of tinnitus (with a higher proportion of TW patients reporting it than in the NE group), multiple events in which participants reported ear fullness : after physical activity (with a higher proportion of TW patients reporting it than in the NE group), after a bad sleep (with a small effect size, a higher proportion of TW and TI patients reporting it than in the NE group), after working at a desk (with a higher proportion of TI patients reporting it than in the NE group), after periods of stress (with a small effect size, a higher proportion of TW and TI patients reporting it than in the NE group); the frequency of occurrence of fleeting tinnitus (with a higher proportion of TW and TI patients reporting it than in the NE group); mixture of tones (with a higher proportion of TW patients reporting it than in the NE group) and electric type of noise (with a higher proportion of TW patients reporting it than in the NE group) as tinnitus type of sounds (not tested in the TRI database); influence of anxiety on tinnitus (with a small effect size, a higher proportion of TW and TI patients reporting a worsening of tinnitus with anxiety compared to the NE group); and influence of physical activity for all intensities from light exercise to intense workout (in all cases, more NE group responders reported physical activity had no effect on their tinnitus than in TW and TI).

4. Discussion

Both databases showed that the majority of participants did not experience any influence on tinnitus by a nap. The sample characteristics of both databases on this criterion were similar to the ones reported in [12] on another large tinnitus sample (TW: 16.4%, NE: 70.7%, TI: 12.9%) and in [31] on a smaller sample (TW: 23.5%, NE: 64.7%, TI: 11.8%). This could be indicative of a subphenotype with influence of nap on tinnitus, which is found in about 30% of the burdened people with tinnitus. It is noteworthy that in all three data sets, the proportion experiencing worsening of tinnitus from napping is higher than that experiencing improvement of tinnitus. In general rule, our results suggest that one out of five tinnitus sufferers reports increase of tinnitus after naps, while only one out of ten reports an improvement of tinnitus following a nap. It seems surprising that a potentially restorative measure such as a nap has the opposite effect on tinnitus in a certain proportion of people with tinnitus.

In general, it is known that a nap can have divergent effects on health and performance. Some studies show positive effects of napping such as improved alertness, physical and cognitive performance [32–34]. Others, however, show negative effects of napping such as an association with obesity, chronic diseases and overall mortality [35–37]. Particularly noteworthy in this context is a meta-analysis, the results of which suggest that a nap probably increases the risk of depression [38] . Although the results were quite heterogeneous, which was attributed, among other things, to the characteristics of the subject sample, but also to the nap pattern (e.g., duration, frequency, time), genetic aspects, inflammation and hyperactivity of the HPA axis were discussed as underlying mechanisms for the relationship between naps and the risk of depression. On the other hand, depressive symptoms occur frequently in tinnitus patients in clinical settings with a mean prevalence of 33% [39] and have a significant influence on the level of suffering [40,41], with the extent of depressive symptoms contributing significantly to tinnitus severity [42] and tinnitus severity increasing the likelihood of the presence of depressive symptoms [43]. Accordingly, for a certain proportion of tinnitus sufferers, naps could also increase tinnitus severity by increasing on the long run the risk of depression. It should be emphasized that in the present study the nap was not characterized in more detail in terms of duration, frequency or timing, although according to the available literature these aspects are probably important in terms of benefit or harm to health [44,45]. As such, it is also conceivable that the different effects in the study group are due to different characteristics of naps.

Sample characteristics comparison between databases revealed that they differed on the majority of common items. Such differences between samples collected in a clinical environment and samples collected on an online platform on tinnitus have already been reported in the past [29]. It is reassuring to note that the specificities found for the TW, NE and TI populations were vastly congruent between databases despite the general sample characteristics were found significantly different between the two databases (as shown in Table 1). Notably, such specificities were congruent on age, gender, somatosensory maneuvers, initial cause of onset linked to a stressful situation and influence of sound on tinnitus (worsening and suppressing). . Having such congruent findings in two databases collected in two different contexts (online community and clinical environment) and with initial sample differences reflects high ecological validity and robustness of our findings.

In both databases, the largest effect sizes were shown in the group comparison for sleep variables. From this it can be deduced that in people whose tinnitus is modulated by a nap, night sleep is also often an influencing factor on tinnitus. Our data also show that a considerable proportion of tinnitus patients do not experience any influence of night sleep on their tinnitus. While in the TRI database only an influence of night sleep on tinnitus was asked, in the Tinnitus Hub survey database a differentiation of the influence of good or poor night sleep was made, whereby significant group differences were found. Remarkably, in the group TW 33.9% stated to experience a worsening of tinnitus by good night sleep. In NE this was only 6.5% and in TI only 4.7%. In parallel, in TW, the group experiencing worsening of tinnitus due to poor sleep was the largest with 80.8% compared to 43.9% in group NE and 74.8% in group TI, respectively. This implies that tinnitus in group TW is particularly sensitive to change. It is possible that the response pattern of group TW also reflects the clinically known phenomenon for night sleep referred to by "morning roar". This means that tinnitus patients perceive the tinnitus in the morning after waking up as particularly loud and stressful, which is often attributed by the affected persons to the night sleep [46]. Nevertheless, 55.2% of the NE group stated that even poor sleep had no effect on the tinnitus, which could be an expression of a certain resilience to negative influences.

Sleep disturbances have long been considered a symptom of tinnitus, with disturbed sleep being associated with higher tinnitus burden [47–53]. Additionally, some sleep disorders, such as sleep apnea, snoring and night terrors have been reported to be highly co-prevalent with tinnitus [54–57]. More modern concepts postulate a bidirectional relationship between sleep disturbances and higher tinnitus burden, analogous to other disorders such as PTSD or depression [58,59]. In this context, sleep satisfaction represents an important marker for tinnitus burden but also for global health [60]. Different studies show that sleep interventions can improve tinnitus and tinnitus interventions improve sleep [61–65]. However, to date, there is only little evidence on how sleep modulates tinnitus [18,66]. The available data suggest that sleep can also have a negative effect on tinnitus and, in principle, opposite effects such as good night sleep leading to worsening of tinnitus are possible. At the same time, the data suggest that only a subset of people with tinnitus experience modulation by sleep, which could be an expression of general resilience in the NE group or increased sensitivity to change by external influences in the others. Alternatively, specific mechanisms in sleep could have an influence on tinnitus analogous to conditions such as PTSD, in which specific mechanisms, such as REM sleep fragmentation, contribute significantly to the development and maintenance of the disorder [58]. For further clarification, polysomnographic examinations with correlation of tinnitus and sleep recordings over several days would be necessary, which is costly and would require great effort. Only a case report of such an attempt has been reported to date in the appendix of [18]. Reported findings are a correlation between Rapid Eye Movement (REM) sleep duration and overnight tinnitus modulation, as well as a correlation between snoring amount during the night and overnight tinnitus modulation.

In both databases, the influence of distress on tinnitus was more amplified in the TW and TI compared to NE. For many years, the co-activation of non-specific stress networks has been considered to play an important role in the development of tinnitus distress [67,68]. On the other hand, stress-induced mechanisms such as disrupted cortical networks, dysregulation in the autonomic nervous system, and changes in central serotonin and the HPA axis may play an important role in the development of sleep disorders [69–71]. Reactivity concepts in terms of vulnerability models have been developed for both sleep and stress [72,73]. This means the extent to which an individual reacts to stress exposure with stress reactions or disturbed sleep. An important link is hyperarousal, which plays a significant role in both tinnitus and insomnia [74]. Hyperarousal is related to increased activation of the sympathetic nervous system and accompanied by enhanced activity in limbic and autonomic areas of the brain [75]. Overall, it can be postulated that the influence of stress on the interplay between tinnitus and sleep is modulated via the individual sleep or stress reactivity outlined above, and the absence of influence of stress on tinnitus in a subgroup could be an expression of increased resilience.

The more frequent influence of stress on tinnitus in TW and TI groups could tend to suggest specific influence of serotoninergic drive on central auditory circuits supporting tinnitus. In fact, serotonergic excitatory drive from the Dorsal Raphe Nucleus (DRN) has already been suggested as a potential modulator of tinnitus [67,76,77]. Some studies suggest that the increased prevalence of depression in the tinnitus population might be linked to alterations in the serotonergic system [77,78]. Since then, some studies analyzed serotonin characteristics in the tinnitus population, and found diverging results both for blood (hence peripheral) serotonin measurements [79–81] and central serotonin correlates [82–84]. Yet, these studies made their measurements on the whole tinnitus population without accounting for its intrinsic heterogeneity [85], hence what has not been revealed to hold true for the whole tinnitus population may be true for some of its subgroups, such as TW or TI. In fact, it is well-known that serotonergic activity fluctuates between wake and sleep and within sleep stages, notably with REM [86–89]. Notably, such modulations have also been reported to happen sometimes during naps [90]. An additional congruent interesting observation is that physical activity also modulates specifically more tinnitus in TI and TW populations. It has been reported in past studies that the serotonergic system is particularly involved during physical activity and fatigue following physical activity [91,92]. Such a possible modulation of tinnitus by the serotonergic system should be more closely investigated in future clinical studies. Serotonin is already one of the biomarkers that has been repeatedly studied in relation to prognosis and severity of tinnitus. Nevertheless, the exact role of serotonin in tinnitus has not yet been clarified [93].

Another finding in both databases is the greater proportion of responders experiencing somatosensory modulations of their tinnitus in the TW group. Such observation suggests a potential link between these phenomena, as anecdotally suggested in past reviews [16–18]. A potential explanation of this association could be that increases in tinnitus following naps could actually be somatosensory increases of tinnitus following muscular events occurring during the nap, only perceived upon awakening. One could hypothesize that the TW population often experience bruxism events during naps, thus sending nociceptive signals from the trigeminal-innervated masticatory muscles to the spinal trigeminal nucleus, which would then in turn excite the cochlear nuclei, as described in [94]. It is important to note that such a scenario is yet to be confirmed as the TW group did not show higher jaw pain complaints compared to the other groups in the Tinnitus Hub survey database, and bruxism and jaw pain complaints have shown to be minimally correlated in the general population [95,96]. Another possibility would be that tinnitus sufferers from the TW population often experience sleep apnea and/or snoring events during their sleep. Such events most often involve the soft palate zone and most specifically the Tensor Veli Palatini muscle (TVP) [97], which is another muscle innervated by the trigeminal nerve. It has been reported that a proliferation of nociceptors in the soft palate occur in case of chronic sleep apnea and snoring occurrences [98], as well as an increase of density of the noradrenergic terminals in the trigeminal sensory nuclei in animals subjected to chronic intermittent hypoxia [99]. Such hypersensibility could increase the risk of somatosensory modulations. Repeated activations of the TVP caused by apnea or snoring events during naps could then trigger a somatosensory modulation of tinnitus upon awakening. A correlation between overnight duration of snoring and overnight modulation of tinnitus has been reported in the 7-night polysomnography longitudinal case report in appendix of [18]. These hypotheses are worth further future investigation and should be tested by polysomnographic or polygraphic investigations.

A set of specific observations in the TI population may highlight a possible explanation to the improvements they report following naps. In fact, we observe that TI patients have a higher prevalence of hyperacusis symptoms than the TW and NE groups (in the TRI database), they also report more often pain and/or discomfort when exposed to sounds and their tinnitus worsens with sound exposition. From these observations, it could be hypothesized that by taking a nap in a calm and quiet environment it would appear logical that such patients report improvements in their symptoms.

On the other hand, it is remarkable to observe that regarding the TW population, there are a lot of factors that seem to be able to increase tinnitus : stress, sleep, physical activity, sound exposition and somatosensory modulations. Such observation tends to highlight a global hypersensitivity of a subgroup of tinnitus patients to external stimuli.

Limitations:

It should be noted that for common items in both databases, wording of the questions and associated possible answers often varied between databases. Even if best possible matching was done to prevent discrepancies, part of divergences between the characteristics of the TW, NE and TI populations between databases may at least partly stem from these wording differences. However, the largely consistent results from the two databases in spite of differences in population characteristics and wording of questions underscores the robustness of our findings

Besides, by comparing the average scores of each group for all questionnaire items, an implicit assumption was made that there were no distinct subgroups of patients within the TI, TW and NE populations. Specific additional analyses should be conducted to study whether or not there is heterogeneity within the TI, TW and NE groups. Such analyses could take the form of a cluster analysis and may bring more insights on the question of whether TW and/or TI groups are composed of patients who developed tinnitus due to only one or several different pathophysiological mechanisms.

## 5. Conclusions

In this retrospective comparative study of two large samples with 9,742 participants, the influence of napping on tinnitus was systematically investigated for the first time. Consistent with previous data, only a minority reported that tinnitus could be influenced by naps, with significantly more in this group reporting a negative than a positive influence on tinnitus.

The influence of napping on tinnitus was associated in both databases with the modulation of tinnitus intensity by night sleep quality, by stress, by somatosensory maneuvers and physical activity. Group TI showed an association with sound sensitivity, hyperacusis and pain, suggesting that taking a nap combined with resting in a calm and quiet environment may improve symptoms. The TW group showed an association with general hypersensitivity to tinnitus, which may be modulated by multiple factors such as distress, sleep, physical activity, noise exposure or somatosensory modulation. Possible underlying mechanisms of nap modulation on tinnitus were discussed but cannot be conclusively clarified.

Although the two groups were recruited very differently, very similar results were found regarding group distribution and the association of other influencing factors. Accordingly, the risk of a sample bias is minimized and the findings can be considered as robust.

Future studies should use polysomnography to investigate the phenomenon of nap modulation in tinnitus patients to better understand the underlying mechanisms and also the divergent effect. In this context, the potential role of the central serotonin system in modulation should also be investigated, as well as the complex interaction of somatosensory and nap modulation on tinnitus.

Likewise, further research may focus on the potential heterogeneity within each of these groups.

**Supplementary Material:** Joined to this original article are Supplementary Tables 1 and 2, which consists in extended versions of Tables 2 and 3 where variables that did not show differences between the three groups TW, NE and TI are included.

**Author Contributions:** Conceptualization, R.G., M.S., J.S., B.L. and F.C.W; methodology, R.G., M.S., J.S., B.L. and F.C.W; software, R.G. and J.S.; validation, R.G., M.S., J.S., B.L. and F.C.W; formal analysis, R.G., M.S., J.S., B.L. and F.C.W; investigation, R.G., M.S., J.S., B.L. and F.C.W; resources, R.G., S.M., M.V., H.G., M.S., J.S., B.L. and F.C.W.; data curation, S.M., R.G and J.S.; writing—original draft preparation, R.G. and F.C.W; writing—review and editing, R.G., M.S., J.S., B.L., A.L., M.C, S.M., M.V., H.G., T.W. and F.C.W; visualization, R.G.; supervision, B.L, M.S, A.L., M.C., SM. and F.C.W.; funding acquisition, R.G. All authors have read and agreed to the published version of the manuscript.

**Funding:** This research was funded by Fondation Félicia et Jean-Jacques Lopez-Loreta.

**Institutional Review Board Statement:** The study was conducted in accordance with the Declaration of Helsinki, and approved for the Tinnitus Hub survey database by the Ethics Committee of the Antwerp University Hospital, reference number : 19-43-485. It was approved for the TRI database by the Ethics Committee of the University of Regensburg, protocol number 08/046.

**Informed Consent Statement:** Informed consent was obtained from all subjects involved in the study.

**Data Availability Statement:** The code used to perform this analysis is accessible on the GitHub open-source repository at this URL : https://github.com/RobinGuillard/TinniNap_DB_study The TRI data presented in this study are available on reasonable request from the corresponding author. The Tinnitus Hub survey database is available on reasonable request to Tinnitus Hub and S.M.

**Acknowledgements:** we wanted to warmly thank Mrs Susanne Staudinger for her help and assistance in handling the TRI database.

**Conflicts of Interest:** Robin Guillard declares that he is a shareholder and president in Siopi SAS and has a professional activity as independent as Robin Guillard EIRL. The other authors did not declare Conflicts Of Interest. The funders had no role in the design of the study; in the collection, analyses, or interpretation of data; in the writing of the manuscript; or in the decision to publish the results.

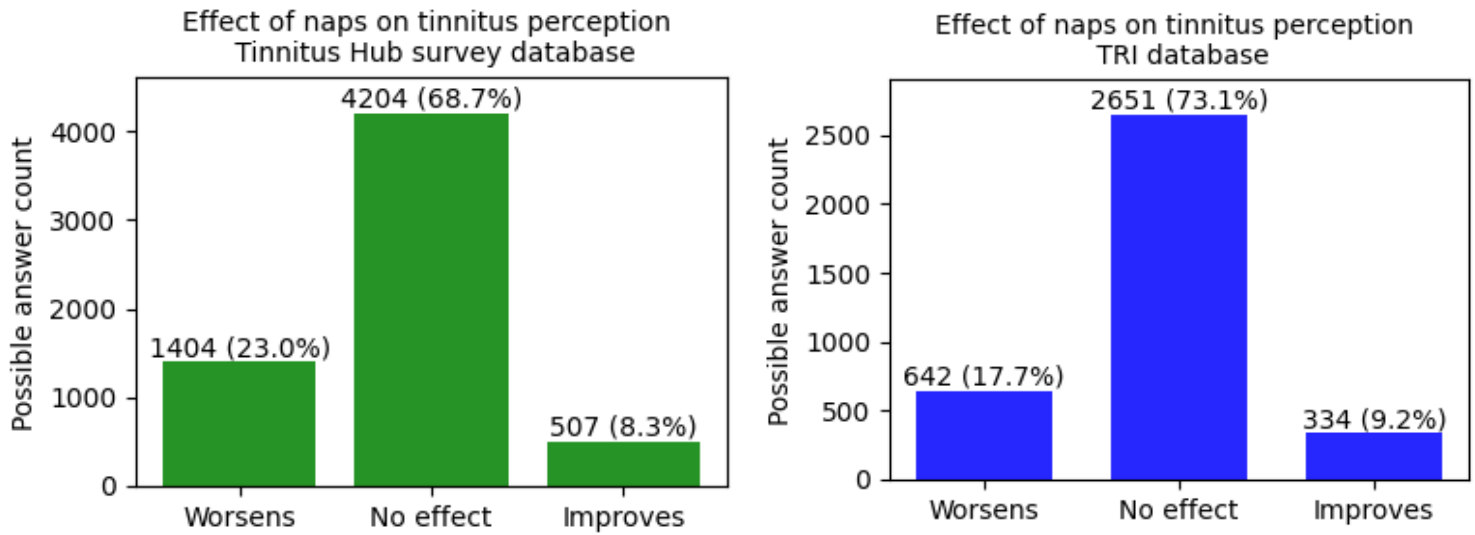

*Figure 1 left : Repartition of the sample between groups on the base of the effect of naps on tinnitus percept for the Tinnitus Hub survey database*

*Figure 2 right : Repartition of the sample between groups on the base of the effect of naps on tinnitus perception for the TRI Database*

**Table 1**. Socio-demographic and tinnitus characteristics in each database and comparison of such characteristics. *SD: Standard Deviation, *: significant p-value*

|  | TRI database (N= 3627) | Tinnitus Hub database (N= 6115) | Statistic | P-Value | Effect size |
|---|---|---|---|---|---|
| **Age** |  |  | H = 13465652* | p < 0.001 | 0.215 (Small) |
| Mean (SD) | 59.8 (13.85) | 54.1 (13.81) |  |  |  |
| Median [Min Max] | 61.0 [2.0, 103.0] | 57.0 [6, 96] |  |  |  |
| Missing | 1 (0.0%) | 0 (0.0%) |  |  |  |
| **Loudness** |  |  | H = 12739120* | p < 0.001 | 0.205 (Small) |
| Mean (SD) | 6.4 (2.19) | 5.6 (2.33) |  |  |  |
| Median [Min Max] | 7.0 [0.0, 10.0] | 6.0 [0.0, 10.0] |  |  |  |
| Missing | 156 (4.3%) | 24 (0.4%) |  |  |  |
| **Annoyance** |  |  | H = 13115675* | p < 0.001 | 0.239 (Small) |
| Mean (SD) | 6.8 (2.43) | 5.5 (3.04) |  |  |  |
| Median [Min Max] | 7.0 [0.0, 10.0] | 5.0 [0.0, 10.0] |  |  |  |
| Missing | 152 (4.2%) | 22 (0.4%) |  |  |  |
| **Gender** |  |  | Chi2 = 186.8* | p < 0.001 | 0.139 (Small) |
| Male | 2392 (65.9%) | 3154 (51.7%) |  |  |  |
| Female | 1235 (34.1%) | 2942 (48.3%) |  |  |  |
| Missing | 0 (0.0%) | 19 (0.3%) |  |  |  |
| **Initial cause stress** |  |  | Chi2 = 1136.3* | p < 0.001 | 0.352 (Medium) |
| No | 1508 (49.4%) | 5076 (83.0%) | | | |
| Yes | 1546 (50.6%) | 1039 (17.0%) | | | |
| Missing | 573 (15.8%) | 0 (0.0%) | | | |
| **Initial cause acoustic trauma** | | | Chi2 = 312.2* | p < 0.001 | 0.185 (Small) |
| No | 2950 (96.6%) | 5131 (83.9%) | | | |
| Yes | 104 (3.4%) | 984 (16.1%) | | | |
| Missing | 573 (15.8%) | 0 (0.0%) | | | |
| **Hearing impairment** | | | Chi2 = 426.0* | p < 0.001 | 0.210 (Small) |
| No | 1401 (39.3%) | 1219 (19.9%) | | | |
| Yes | 2166 (60.7%) | 4896 (80.1%) | | | |
| Missing | 60 (1.7%) | 0 (0.0%) | | | |
| **Somatosensory modulations** | | | Chi2 = 340.1* | p < 0.001 | 0.187 (Small) |
| No | 2258 (63.2%) | 2678 (43.8%) | | | |
| Yes | 1313 (36.8%) | 3437 (56.2%) | | | |
| Missing | 56 (1.5%) | 0 (0.0%) | | | |
| **Some sounds worsen tinnitus** | | | Chi2 = 349.7* | p < 0.001 | 0.197 (Small) |
| No | 916 (31.2%) | 3185 (52.1%) | | | |
| Yes | 2023 (68.8%) | 2930 (47.9%) | | | |
| Missing | 688 (19.0%) | 0 (0.0%) | | | |
| **Some sounds suppress tinnitus** | | | Chi2 = 334.4* | p < 0.001 | 0.191 (Small) |
| No | 723 (23.7%) | 580 (9.5%) | | | |
| Yes | 2334 (76.3%) | 5535 (90.5%) | | | |
| Missing | 570 (15.7%) | 0 (0.0%) | | | |

*Table 1 : Socio-demographic and tinnitus characteristics in each database and comparison of such characteristics. A quantification specific to the effect size calculated is systematically given. SD : Standard Deviation*

**Table 2**. Group comparison in the TRI database. Only variables eliciting significant differences after Holm correction were reported. For comparison on the complete list of variables, please refer to Supplementary material 1. *SD : Standard Deviation, TW : Tinnitus worsens after naps, NE : No effect of naps on tinnitus, TI : Tinnitus improves after naps, THI : Tinnitus Handicap Inventory, TQ : Tinnitus Questionnaire, \* : p < 0.05 before Holm correction, \*\* : p < 0.05 after Holm correction.*

|  | Worsens N= 642 | No effect N= 2651 | Improves N= 334 | Statistic | p-value | Effect size | Post-hoc | Post-hoc p-value | Post-hoc effect size |
|---|---|---|---|---|---|---|---|---|---|
| **THI score (0-100)** |  |  |  | H = 53.8\*\* | p < 0.001 | 0.015 (Small) |  |  |  |
| Mean (SD) | 53.6 (22.29) | 46.8 (23.33) | 52.2 (21.75) |  |  |  | TW VS NE | p < 0.001 | 0.297 (Small) |
| Median [Min Max] | 54.0 [0.0, 100] | 46.0 [0, 100] | 50.0 [4, 100] |  |  |  | TI VS NE | 0.002 | 0.236 (Small) |
| Missing | 13 (2.0%) | 41 (1.5%) | 6 (1.8%) |  |  |  | TW VS TI | 1.0 | 0.064 (Negligible) |
| **Tinnitus varies from day to day** |  |  |  | Chi2 = 136.9\*\* | p < 0.001 | 0.137 (Small) |  |  |  |
| Yes | 489 (77.1%) | 1485 (56.8%) | 264 (80.0%) |  |  |  | TW VS NE | p < 0.001 | 0.155 (Small) |
| No | 145 (22.9%) | 1128 (43.2%) | 66 (20.0%) |  |  |  | TI VS NE | p < 0.001 | 0.133 (Small) |
| Missing | 8 (1.2%) | 38 (1.4%) | 4 (1.2%) |  |  |  | TW VS TI | 1.0 | 0.016 (Negligible) |
| **Tinnitus intermittent or continuous ?** |  |  |  | Chi2 = 38.9\*\* | p < 0.001 | 0.073 (Small) |  |  |  |
| Intermittent | 112 (17.8%) | 307 (11.7%) | 74 (22.4%) |  |  |  | TW VS NE | 0.002 | 0.067 (Negligible) |
| Continuous | 517 (82.2%) | 2317 (88.3%) | 256 (77.6%) |  |  |  | TI VS NE | p < 0.001 | 0.09 (Negligible) |
| Missing | 13 (2.0%) | 27 (1.0%) | 4 (1.2%) |  |  |  | TW VS TI | 1.0 | 0.027 (Negligible) |
| **Influence of stress over tinnitus** |  |  |  | Chi2 = 84.2\*\* | p < 0.001 | 0.076 (Small) |  |  |  |
| Worsens | 477 (77.2%) | 1742 (67.5%) | 294 (89.6%) |  |  |  | TW VS NE | p < 0.001 | 0.057 (Negligible) |
| Improves | 8 (1.3%) | 29 (1.1%) | 0 (0.0%) |  |  |  | TI VS NE | p < 0.001 | 0.097 (Small) |
| No effect | 133 (21.5%) | 809 (31.4%) | 34 (10.4%) |  |  |  | TW VS TI | p < 0.001 | 0.057 (Negligible) |
| Missing | 24 (3.7%) | 71 (2.7%) | 6 (1.8%) |  |  |  |  |  |  |
| **Sleep at night and tinnitus during the day** |  |  |  | Chi2 = 399.0\*\* | p < 0.001 | 0.235 (Medium) |  |  |  |
| Linked | 261 (71.5%) | 338 (22.8%) | 127 (68.6%) |  |  |  | TW VS NE | p < 0.001 | 0.295 (Small) |
| Not linked | 104 (28.5%) | 1145 (77.2%) | 58 (31.4%) |  |  |  | TI VS NE | p < 0.001 | 0.216 (Small) |
| Missing | 277 (43.1%) | 1168 (44.1%) | 149 (44.6%) |  |  |  | TW VS TI | 1.0 | 0.01 (Negligible) |
| **Somatosensory : Jaw or head movements can modulate tinnitus** |  |  |  | Chi2 = 98.6\*\* | p < 0.001 | 0.117 (Small) |  |  |  |
| Yes | 325 (51.5%) | 835 (32.0%) | 153 (46.6%) |  |  |  | TW VS NE | p < 0.001 | 0.152 (Small) |
| No | 306 (48.5%) | 1777 (68.0%) | 175 (53.4%) |  |  |  | TI VS NE | p < 0.001 | 0.087 (Negligible) |
| Missing | 11 (1.7%) | 39 (1.5%) | 6 (1.8%) |  |  |  | TW VS TI | 1.0 | 0.023 (Negligible) |

*Table 2 : Group comparison in the TRI database. Only variables eliciting significant differences after Holm correction were reported. For comparison on the complete list of variables, please refer to Supplementary material 1. A quantification specific to the effect size calculated is systematically given. SD : Standard Deviation*

**Table 3**. Group comparison in the Tinnitus Hub survey database. Only variables eliciting significant differences after Holm correction were reported. For comparison on the complete list of variables, please refer to Supplementary Table 2. *SD : Standard Deviation, TW : Tinnitus worsens after naps, NE : No effect of naps on tinnitus, TI : Tinnitus improves after naps, \* : p < 0.05 before Holm correction, \*\* : p < 0.05 after Holm correction.*

| | Worsens N= 1404 | No effect N= 4204 | Improves N= 507 | Statistic | p-value | Effect size | Post-hoc test | | Post-hoc effect size |
|---|---|---|---|---|---|---|---|---|---|
| **Gender** | | | | Chi2 = 59.6** | p < 0.001 | 0.07 (Small) | | | |
| Male | 832 (59.7%) | 2031 (48.6%) | 291 (57.9%) | | | | TW VS NE | p < 0.001 | 0.092 (Negligible) |
| Female | 562 (40.3%) | 2151 (51.4%) | 212 (42.1%) | | | | TI VS NE | 0.005 | 0.05 (Negligible) |
| Missing | 10 (0.7%) | 22 (0.5%) | 4 (0.8%) | | | | TW VS TI | 1.0 | 0.008 (Negligible) |
| **Tinnitus cause : psychological (stress, anxiety, depression)** | | | | Chi2 = 60.1** | p < 0.001 | 0.07 (Small) | | | |
| No | 1101 (78.4%) | 3593 (85.5%) | 382 (75.3%) | | | | TW VS NE | p < 0.001 | 0.079 (Negligible) |
| Yes | 303 (21.6%) | 611 (14.5%) | 125 (24.7%) | | | | TI VS NE | p < 0.001 | 0.075 (Negligible) |
| Missing | 0 (0.0%) | 0 (0.0%) | 0 (0.0%) | | | | TW VS TI | 1.0 | 0.017 (Negligible) |
| **Fluctuations of tinnitus** | | | | Chi2 = 257.3** | p < 0.001 | 0.084 (Small) | | | |
| No fluctuations | 268 (19.1%) | 1512 (36.0%) | 80 (15.8%) | | | | TW VS NE | p < 0.001 | 0.092 (Small) |
| Grows louder as day progresses | 703 (50.1%) | 1782 (42.4%) | 218 (43.0%) | | | | TI VS NE | p < 0.001 | 0.084 (Small) |
| Grows quieter as day progresses | 122 (8.7%) | 193 (4.6%) | 36 (7.1%) | | | | TW VS TI | p < 0.001 | 0.039 (Negligible) |
| Changes within the day or over days without any pattern | 311 (22.2%) | 717 (17.1%) | 173 (34.1%) | | | | | | |
| Missing | 0 (0.0%) | 0 (0.0%) | 0 (0.0%) | | | | | | |
| **Influence of stress over tinnitus** | | | | Chi2 = 292.6** | p < 0.001 | 0.109 (Small) | | | |
| Worsens | 1128 (80.3%) | 2439 (58.0%) | 409 (80.7%) | | | | TW VS NE | p < 0.001 | 0.136 (Small) |
| No effect | 271 (19.3%) | 1746 (41.5%) | 95 (18.7%) | | | | TI VS NE | p < 0.001 | 0.09 (Small) |
| Improves | 5 (0.4%) | 19 (0.5%) | 3 (0.6%) | | | | TW VS TI | 1.0 | 0.007 (Negligible) |
| Missing | 0 (0.0%) | 0 (0.0%) | 0 (0.0%) | | | | | | |
| **Influence of anxiety over tinnitus** | | | | Chi2 = 317.5** | p < 0.001 | 0.114 (Small) | | | |
| Worsens | 1090 (77.6%) | 2272 (54.0%) | 394 (77.7%) | | | | TW VS NE | p < 0.001 | 0.141 (Small) |
| No effect | 309 (22.0%) | 1910 (45.4%) | 107 (21.1%) | | | | TI VS NE | p < 0.001 | 0.095 (Small) |
| Improves | 5 (0.4%) | 22 (0.5%) | 6 (1.2%) | | | | TW VS TI | 1.0 | 0.019 (Negligible) |
| Missing | 0 (0.0%) | 0 (0.0%) | 0 (0.0%) | | | | | | |
| **Influence of a good night sleep over tinnitus** | | | | Chi2 = 1373.5** | p < 0.001 | 0.237 (Medium) | | | |
| Worsens | 476 (33.9%) | 272 (6.5%) | 24 (4.7%) | | | | TW VS NE | p < 0.001 | 0.257 (Medium) |
| No effect | 430 (30.6%) | 2665 (63.4%) | 67 (13.2%) | | | | TI VS NE | p < 0.001 | 0.21 (Medium) |
| Improves | 498 (35.5%) | 1267 (30.1%) | 416 (82.1%) | | | | TW VS TI | p < 0.001 | 0.165 (Small) |
| Missing | 0 (0.0%) | 0 (0.0%) | 0 (0.0%) | | | | | | |
| **Influence of poor sleep over tinnitus** | | | | Chi2 = 694.5** | p < 0.001 | 0.169 (Medium) | | | |
| Worsens | 1134 (80.8%) | 1844 (43.9%) | 379 (74.8%) | | | | TW VS NE | p < 0.001 | 0.22 (Medium) |
| No effect | 250 (17.8%) | 2320 (55.2%) | 118 (23.3%) | | | | TI VS NE | p < 0.001 | 0.123 (Small) |
| Improves | 20 (1.4%) | 40 (1.0%) | 10 (2.0%) | | | | TW VS TI | 0.706 | 0.026 (Negligible) |
| Missing | 0 (0.0%) | 0 (0.0%) | 0 (0.0%) | | | | | | |
| **Tinnitus sound masking** | | | | Chi2 = 128.7** | p < 0.001 | 0.046 (Small) | | | |
| No masking | 101 (7.2%) | 410 (9.8%) | 69 (13.6%) | | | | TW VS NE | p < 0.001 | 0.058 (Small) |
| Only a small selection of specific sounds | 162 (11.5%) | 357 (8.5%) | 63 (12.4%) | | | | TI VS NE | p < 0.001 | 0.033 (Negligible) |
| Shower / water noises | 287 (20.4%) | 485 (11.5%) | 68 (13.4%) | | | | TW VS TI | p < 0.001 | 0.033 (Negligible) |
| Television, Music or general background noise | 334 (23.8%) | 1291 (30.7%) | 135 (26.6%) | | | | | | |
| White noise or special masking noises | 197 (14.0%) | 555 (13.2%) | 80 (15.8%) | | | | | | |
| Masked by nearly all sounds | 323 (23.0%) | 1106 (26.3%) | 92 (18.1%) | | | | | | |
| Missing | 0 (0.0%) | 0 (0.0%) | 0 (0.0%) | | | | | | |
| **Influence of intense workout over tinnitus** | | | | Chi2 = 252.2** | p < 0.001 | 0.102 (Small) | | | |
| Worsens | 424 (30.2%) | 645 (15.3%) | 126 (24.9%) | | | | TW VS NE | p < 0.001 | 0.121 (Small) |
| No effect | 847 (60.3%) | 3282 (78.1%) | 295 (58.2%) | | | | TI VS NE | p < 0.001 | 0.096 (Small) |
| Improves | 133 (9.5%) | 277 (6.6%) | 86 (17.0%) | | | | TW VS TI | p < 0.001 | 0.043 (Negligible) |
| Missing | 0 (0.0%) | 0 (0.0%) | 0 (0.0%) | | | | | | |
| **Influence of moderate workout over tinnitus** | | | | Chi2 = 326.7** | p < 0.001 | 0.116 (Small) | | | |
| Worsens | 423 (30.1%) | 584 (13.9%) | 117 (23.1%) | | | | TW VS NE | p < 0.001 | 0.135 (Small) |
| No effect | 801 (57.1%) | 3219 (76.6%) | 267 (52.7%) | | | | TI VS NE | p < 0.001 | 0.11 (Small) |
| Improves | 180 (12.8%) | 401 (9.5%) | 123 (24.3%) | | | | TW VS TI | p < 0.001 | 0.056 (Negligible) |
| Missing | 0 (0.0%) | 0 (0.0%) | 0 (0.0%) | | | | | | |
| **Influence of light exercise over tinnitus** | | | | Chi2 = 320.8** | p < 0.001 | 0.115 (Small) | | | |
| Worsens | 231 (16.5%) | 284 (6.8%) | 59 (11.6%) | | | | TW VS NE | p < 0.001 | 0.113 (Small) |
| No effect | 937 (66.7%) | 3424 (81.4%) | 278 (54.8%) | | | | TI VS NE | p < 0.001 | 0.13 (Small) |
| Improves | 236 (16.8%) | 496 (11.8%) | 170 (33.5%) | | | | TW VS TI | p < 0.001 | 0.072 (Small) |
| Missing | 0 (0.0%) | 0 (0.0%) | 0 (0.0%) | | | | | | |
| **Somatosensory change : no change with any of these actions** | | | | Chi2 = 62.4** | p < 0.001 | 0.071 (Small) | | | |
| No | 734 (52.3%) | 1703 (40.5%) | 241 (47.5%) | | | | TW VS NE | p < 0.001 | 0.098 (Negligible) |
| Yes | 670 (47.7%) | 2501 (59.5%) | 266 (52.5%) | | | | TI VS NE | 0.132 | 0.038 (Negligible) |
| Missing | 0 (0.0%) | 0 (0.0%) | 0 (0.0%) | | | | TW VS TI | 1.0 | 0.023 (Negligible) |
| **Ear fullness : after a bad sleep** | | | | Chi2 = 72.7** | p < 0.001 | 0.077 (Small) | | | |
| No | 1250 (89.0%) | 3994 (95.0%) | 452 (89.2%) | | | | TW VS NE | p < 0.001 | 0.1 (Small) |
| Yes | 154 (11.0%) | 210 (5.0%) | 55 (10.8%) | | | | TI VS NE | p < 0.001 | 0.068 (Negligible) |
| Missing | 0 (0.0%) | 0 (0.0%) | 0 (0.0%) | | | | TW VS TI | 1.0 | 0.0 (Negligible) |
| **Ear fullness : after periods of stress / anxiety** | | | | Chi2 = 85.3** | p < 0.001 | 0.084 (Small) | | | |
| No | 1129 (80.4%) | 3700 (88.0%) | 386 (76.1%) | | | | TW VS NE | p < 0.001 | 0.091 (Negligible) |
| Yes | 275 (19.6%) | 504 (12.0%) | 121 (23.9%) | | | | TI VS NE | p < 0.001 | 0.094 (Negligible) |
| Missing | 0 (0.0%) | 0 (0.0%) | 0 (0.0%) | | | | TW VS TI | 1.0 | 0.025 (Negligible) |

*Table 3 : Group comparison in the Tinnitus Hub survey database. Only variables eliciting significant differences after Holm correction were reported. For comparison on the complete list of variables, please refer to Supplementary material 2. A quantification specific to the effect size calculated is systematically given. SD : Standard Deviation*

**Supplementary Table 1**. Extended group comparison in the TRI database. *SD : Standard Deviation, TW : Tinnitus worsens after naps, NE : No effect of naps on tinnitus, TI : Tinnitus improves after naps, THI : Tinnitus Handicap Inventory, TQ : Tinnitus Questionnaire, WHOQOL : The World Health Organization Quality of Life, * : p < 0.05 before Holm correction, ** : p < 0.05 after Holm correction.*

|  | Worsens N= 642 | No effect N= 2651 | Improves N= 334 | Statistic | p-value | Effect size | Post-hoc test |  | Post-hoc effect size |
|---|---|---|---|---|---|---|---|---|---|
| **Age (years)** |  |  |  | H = 13.2** | 0.04 | 0.003 (Negligible) |  |  |  |
| Mean (SD) | 61.8 (14.67) | 59.4 (13.45) | 59.7 (14.6) |  |  |  | TW VS NE | 0.012 | 0.175 (Negligible) |
| Median [Min Max] | 61.5 [21, 102] | 60 [18, 103] | 60 [21, 100] |  |  |  | TI VS NE | 1.0 | 0.024 (Negligible) |
| Missing | 0 (0%) | 2 (0.1%) | 0 (0%) |  |  |  | TW VS TI | 0.52 | 0.141 (Negligible) |
| **Tinnitus duration (months)** |  |  |  | H = 2.2 | 1.0 | 0.0 (Negligible) |  |  |  |
| Mean (SD) | 103.9 (111.89) | 100.9 (112.93) | 93.4 (102.16) |  |  |  |  |  |  |
| Median [Min Max] | 55 [0.8, 604] | 53 [0.5, 726] | 49.1 [0.6, 580] |  |  |  |  |  |  |
| Missing | 70 (10.9%) | 270 (10.2%) | 33 (9.9%) |  |  |  |  |  |  |
| **Tinnitus pitch (0: very high frequency, 1: high, 2: medium, 3: low frequency)** |  |  |  | H = 8.3* | 0.248 | 0.002 (Negligible) |  |  |  |
| Mean (SD) | 0.9 (0.72) | 1 (0.74) | 1 (0.77) |  |  |  |  |  |  |
| Median [Min Max] | 1 [0, 3] | 1 [0, 3] | 1 [0, 3] |  |  |  |  |  |  |
| Missing | 35 (5.5%) | 84 (3.2%) | 11 (3.3%) |  |  |  |  |  |  |
| **Tinnitus laterality (0 : bilateral, 0.5 : More on one side, 1 : unilateral)** |  |  |  | H = 6.5* | 0.541 | 0.001 (Negligible) |  |  |  |
| Mean (SD) | 0.5 (0.36) | 0.5 (0.4) | 0.5 (0.4) |  |  |  |  |  |  |
| Median [Min Max] | 0.5 [0, 1] | 0.5 [0, 1] | 0.5 [0, 1] |  |  |  |  |  |  |
| Missing | 5 (0.8%) | 13 (0.5%) | 1 (0.3%) |  |  |  |  |  |  |
| **Average loudness (0-100)** |  |  |  | H = 13.5** | 0.035 | 0.003 (Negligible) |  |  |  |
| Mean (SD) | 66.6 (22.57) | 64.4 (22.63) | 61.1 (23.43) |  |  |  | TW VS NE | 0.52 | 0.098 (Negligible) |
| Median [Min Max] | 70 [0, 100] | 70 [0, 100] | 60 [0, 100] |  |  |  | TI VS NE | 0.305 | 0.145 (Negligible) |
| Missing | 27 (4.2%) | 85 (3.2%) | 7 (2.1%) |  |  |  | TW VS TI | 0.01 | 0.24 (Small) |
| **THI score (0-100)** |  |  |  | H = 53.8** | p < 0.001 | 0.015 (Small) |  |  |  |
| Mean (SD) | 53.6 (22.29) | 46.8 (23.33) | 52.2 (21.75) |  |  |  | TW VS NE | p < 0.001 | 0.297 (Small) |
| Median [Min Max] | 54 [0, 100] | 46 [0, 100] | 50 [4, 100] |  |  |  | TI VS NE | 0.002 | 0.236 (Small) |
| Missing | 13 (2%) | 41 (1.5%) | 6 (1.8%) |  |  |  | TW VS TI | 1.0 | 0.064 (Negligible) |
| **TQ score (0-84)** |  |  |  | H = 23.8** | p < 0.001 | 0.008 (Negligible) |  |  |  |
| Mean (SD) | 43.9 (17.19) | 39.7 (17.91) | 41.5 (16.76) |  |  |  | TW VS NE | p < 0.001 | 0.233 (Small) |
| Median [Min Max] | 44 [4, 84] | 39 [1, 84] | 41 [4, 79] |  |  |  | TI VS NE | 1.0 | 0.099 (Negligible) |
| Missing | 130 (20.2%) | 612 (23.1%) | 75 (22.5%) |  |  |  | TW VS TI | 1.0 | 0.14 (Negligible) |
| **How much of a problem is your tinnitus at present? (0 : no problem, 5 : a very big problem** |  |  |  | H = 18.4** | 0.004 | 0.005 (Negligible) |  |  |  |
| Mean (SD) | 2.6 (0.9) | 2.4 (0.89) | 2.4 (0.85) |  |  |  | TW VS NE | 0.001 | 0.193 (Negligible) |
| Median [Min Max] | 3 [0, 4] | 2 [0, 4] | 2 [0, 4] |  |  |  | TI VS NE | 1.0 | 0.014 (Negligible) |
| Missing | 53 (8.3%) | 187 (7.1%) | 35 (10.5%) |  |  |  | TW VS TI | 0.098 | 0.209 (Small) |
| **How STRONG or LOUD is your tinnitus at present? (0 to 10 scale)** |  |  |  | H = 15.2** | 0.017 | 0.004 (Negligible) |  |  |  |
| Mean (SD) | 6.7 (2.15) | 6.4 (2.19) | 6.1 (2.23) |  |  |  | TW VS NE | 0.353 | 0.108 (Negligible) |
| Median [Min Max] | 7 [0, 10] | 7 [0, 10] | 6 [0, 10] | | | | TI VS NE | 0.235 | 0.154 (Negligible) |
| Missing | 34 (5.3%) | 98 (3.7%) | 24 (7.2%) | | | | TW VS TI | 0.005 | 0.263 (Small) |
| **How UNCOMFORTABLE is your tinnitus at present, if everything around you is quiet? (0 to 10 scale)** | | | | H = 27.9** | p < 0.001 | 0.007 (Negligible) | | | |
| Mean (SD) | 7.2 (2.23) | 7.1 (2.33) | 6.4 (2.47) | | | | TW VS NE | 1.0 | 0.059 (Negligible) |
| Median [Min Max] | 8 [0, 10] | 8 [0, 10] | 7 [1, 10] | | | | TI VS NE | p < 0.001 | 0.303 (Small) |
| Missing | 33 (5.1%) | 104 (3.9%) | 25 (7.5%) | | | | TW VS TI | p < 0.001 | 0.367 (Small) |
| **How ANNOYING is your tinnitus at present? (0 to 10 scale)** | | | | H = 10.6* | 0.111 | 0.002 (Negligible) | | | |
| Mean (SD) | 6.9 (2.41) | 6.8 (2.44) | 6.4 (2.42) | | | | | | |
| Median [Min Max] | 8 [0, 10] | 7 [0, 10] | 7 [0, 10] | | | | | | |
| Missing | 31 (4.8%) | 97 (3.7%) | 24 (7.2%) | | | | | | |
| **How easy is it for you to IGNORE your tinnitus at present? (0 to 10 scale)** | | | | H = 14.1** | 0.026 | 0.003 (Negligible) | | | |
| Mean (SD) | 7.1 (2.51) | 6.9 (2.69) | 6.4 (2.71) | | | | TW VS NE | 1.0 | 0.082 (Negligible) |
| Median [Min Max] | 8 [0, 10] | 7 [0, 10] | 7 [0, 10] | | | | TI VS NE | 0.044 | 0.185 (Negligible) |
| Missing | 30 (4.7%) | 98 (3.7%) | 24 (7.2%) | | | | TW VS TI | 0.008 | 0.277 (Small) |
| **How UNPLEASANT is your tinnitus at present? (0 to 10 scale)** | | | | H = 12.6* | 0.051 | 0.003 (Negligible) | | | |
| Mean (SD) | 7 (2.35) | 6.8 (2.44) | 6.4 (2.41) | | | | | | |
| Median [Min Max] | 8 [0, 10] | 7 [0, 10] | 7 [0, 10] | | | | | | |
| Missing | 30 (4.7%) | 94 (3.5%) | 23 (6.9%) | | | | | | |
| **% of time aware of tinnitus (0-100)** | | | | H = 2.0 | 1.0 | -0.0 (Negligible) | | | |
| Mean (SD) | 73.8 (25.31) | 71.1 (27.98) | 70.7 (27.56) | | | | | | |
| Median [Min Max] | 80 [5, 100] | 80 [0, 100] | 80 [1, 100] | | | | | | |
| Missing | 12 (1.9%) | 20 (0.8%) | 6 (1.8%) | | | | | | |
| **% of time annoyed, distressed, irritated by your tinnitus (0-100)** | | | | H = 10.6* | 0.111 | 0.002 (Negligible) | | | |
| Mean (SD) | 61.5 (29.45) | 56.9 (31.14) | 59.3 (30.77) | | | | | | |
| Median [Min Max] | 70 [0, 100] | 50 [0, 100] | 60 [0, 100] | | | | | | |
| Missing | 9 (1.4%) | 44 (1.7%) | 3 (0.9%) | | | | | | |
| **WHOQOL : Physical Health (0-20)** | | | | H = 9.5* | 0.154 | 0.003 (Negligible) | | | |
| Mean (SD) | 12.4 (1.86) | 12.7 (1.77) | 12.7 (1.75) | | | | | | |
| Median [Min Max] | 13 [7, 18] | 13 [5, 18] | 13 [6, 17] | | | | | | |
| Missing | 113 (17.6%) | 448 (16.9%) | 66 (19.8%) | | | | | | |
| **WHOQOL : Psychological Health (0-20)** | | | | H = 12.3* | 0.056 | 0.003 (Negligible) | | | |
| Mean (SD) | 13.4 (2.11) | 13.8 (2) | 13.9 (1.87) | | | | | | |
| Median [Min Max] | 14 [7, 18] | 14 [6, 19] | 14 [7, 18] | | | | | | |
| Missing | 110 (17.1%) | 448 (16.9%) | 65 (19.5%) | | | | | | |
| **WHOQOL : Social Factors (0-20)** | | | | H = 7.9* | 0.288 | 0.002 (Negligible) | | | |
| Mean (SD) | 14.4 (3.41) | 14.8 (3.27) | 14.4 (3.21) | | | | | | |
| Median [Min Max] | 15 [4, 20] | 16 [4, 20] | 15 [5, 20] | | | | | | |
| Missing | 111 (17.3%) | 449 (16.9%) | 67 (20.1%) | | | | | | |
| **WHOQOL : Environmental Factors (0-20)** | | | | H = 12.4* | 0.054 | 0.003 (Negligible) | | | |
| Mean (SD) | 16.3 (2.43) | 16.6 (2.18) | 16.3 (2.23) | | | | | | |
| Median [Min Max] | 16.5 [6, 20] | 17 [6, 20] | 16 [10, 20] | | | | | | |
| Missing | 112 (17.4%) | 437 (16.5%) | 64 (19.2%) | | | | | | |
| **Degree of hyperacusis (0 : None, 4 : Very important)** | | | | H = 25.5** | p < 0.001 | 0.007 (Negligible) | | | |
| Mean (SD) | 2.3 (1.15) | 2.2 (1.21) | 2.5 (1.2) | | | | TW VS NE | 1.0 | 0.088 (Negligible) |
| Median [Min Max] | 2 [0, 4] | 2 [0, 4] | 2 [0, 4] | | | | TI VS NE | p < 0.001 | 0.291 (Small) |
| Missing | 6 (0.9%) | 16 (0.6%) | 1 (0.3%) | | | | TW VS TI | 0.066 | 0.211 (Small) |
| **Gender** | | | | Chi2 = 10.4* | 0.111 | 0.038 (Negligible) | | | |
| Male | 457 (71.2%) | 1726 (65.1%) | 209 (62.6%) | | | | | | |
| Female | 185 (28.8%) | 925 (34.9%) | 125 (37.4%) | | | | | | |
| Missing | 0 (0%) | 0 (0%) | 0 (0%) | | | | | | |
| **Handedness** | | | | Chi2 = 6.1 | 1.0 | 0.021 (Negligible) | | | |
| Right-handed | 517 (81.2%) | 2205 (83.8%) | 274 (82.3%) | | | | | | |
| Ambidextrous | 47 (7.4%) | 181 (6.9%) | 18 (5.4%) | | | | | | |
| Left-handed | 73 (11.5%) | 245 (9.3%) | 41 (12.3%) | | | | | | |
| Missing | 5 (0.8%) | 20 (0.8%) | 1 (0.3%) | | | | | | |
| **Hearing difficulties** | | | | Chi2 = 6.1* | 0.617 | 0.029 (Negligible) | | | |
| Yes | 357 (56.8%) | 1597 (61.2%) | 212 (64.4%) | | | | | | |
| No | 271 (43.2%) | 1013 (38.8%) | 117 (35.6%) | | | | | | |
| Missing | 14 (2.2%) | 41 (1.5%) | 5 (1.5%) | | | | | | |
| **Tinnitus type of sound** | | | | Chi2 = 18.9* | 0.102 | 0.029 (Negligible) | | | |
| Tonal | 366 (58.6%) | 1728 (66.2%) | 192 (58.5%) | | | | | | |
| Noise | 84 (13.4%) | 298 (11.4%) | 43 (13.1%) | | | | | | |
| Criquets | 107 (17.1%) | 378 (14.5%) | 60 (18.3%) | | | | | | |
| Other | 68 (10.9%) | 207 (7.9%) | 33 (10.1%) | | | | | | |
| Missing | 17 (2.6%) | 40 (1.5%) | 6 (1.8%) | | | | | | |
| **Tinnitus sound pulsatile** | | | | Chi2 = 14.8* | 0.111 | 0.032 (Negligible) | | | |
| Pulsatile following heart beats | 76 (12.2%) | 257 (9.9%) | 42 (12.9%) | | | | | | |
| Pulsatile but not following heart beats | 70 (11.3%) | 227 (8.7%) | 42 (12.9%) | | | | | | |
| Not pulsatile | 475 (76.5%) | 2114 (81.4%) | 242 (74.2%) | | | | | | |
| Missing | 21 (3.3%) | 53 (2%) | 8 (2.4%) | | | | | | |
| **Tinnitus side** | | | | Chi2 = 47.6** | p < 0.001 | 0.033 (Negligible) | | | |
| Right ear | 67 (10.5%) | 340 (12.9%) | 52 (15.6%) | | | | TW VS NE | p < 0.001 | 0.044 (Small) |
| Left ear | 77 (12.1%) | 449 (17%) | 61 (18.3%) | | | | TI VS NE | 1.0 | 0.014 (Negligible) |
| Both ears, worse in left | 167 (26.2%) | 553 (21%) | 67 (20.1%) | | | | TW VS TI | 0.086 | 0.03 (Negligible) |
| Both ears, worse in right | 126 (19.7%) | 428 (16.2%) | 56 (16.8%) | | | | | | |
| Both ears, equally | 109 (17.1%) | 618 (23.4%) | 65 (19.5%) | | | | | | |
| Inside the head | 91 (14.3%) | 250 (9.5%) | 32 (9.6%) | | | | | | |
| Elsewhere | 1 (0.2%) | 1 (0%) | 0 (0%) | | | | | | |
| Missing | 4 (0.6%) | 12 (0.5%) | 1 (0.3%) | | | | | | |
| **Family history of tinnitus complaints** | | | | Chi2 = 0.1 | 1.0 | 0.003 (Negligible) | | | |
| Yes | 148 (23.5%) | 621 (23.9%) | 77 (23.5%) | | | | | | |
| No | 483 (76.5%) | 1976 (76.1%) | 251 (76.5%) | | | | | | |
| Missing | 11 (1.7%) | 54 (2%) | 6 (1.8%) | | | | | | |
| **Tinnitus onset** | | | | Chi2 = 3.2 | 1.0 | 0.021 (Negligible) | | | |
| Gradual | 287 (46.8%) | 1273 (50.4%) | 150 (47.2%) | | | | | | |
| Abrupt | 326 (53.2%) | 1255 (49.6%) | 168 (52.8%) | | | | | | |
| Missing | 29 (4.5%) | 123 (4.6%) | 16 (4.8%) | | | | | | |
| **Tinnitus cause : noise trauma** | | | | Chi2 = 1.0 | 1.0 | 0.012 (Negligible) | | | |
| Yes | 526 (96%) | 2147 (96.8%) | 277 (96.2%) | | | | | | |
| No | 22 (4%) | 71 (3.2%) | 11 (3.8%) | | | | | | |
| Missing | 94 (14.6%) | 433 (16.3%) | 46 (13.8%) | | | | | | |
| **Tinnitus cause : psychological (stress, anxiety, depression)** | | | | Chi2 = 21.7** | $p < 0.001$ | 0.055 (Negligible) | | | |
| Yes | 257 (46.9%) | 1143 (51.5%) | 108 (37.5%) | | | | TW VS NE | 1.0 | 0.031 (Negligible) |
| No | 291 (53.1%) | 1075 (48.5%) | 180 (62.5%) | | | | TI VS NE | $p < 0.001$ | 0.073 (Negligible) |
| Missing | 94 (14.6%) | 433 (16.3%) | 46 (13.8%) | | | | TW VS TI | 0.305 | 0.042 (Negligible) |
| **Tinnitus cause : whiplash** | | | | Chi2 = 11.3* | 0.083 | 0.04 (Negligible) | | | |
| Yes | 510 (93.1%) | 2129 (96%) | 268 (93.1%) | | | | | | |
| No | 38 (6.9%) | 89 (4%) | 20 (6.9%) | | | | | | |
| Missing | 94 (14.6%) | 433 (16.3%) | 46 (13.8%) | | | | | | |
| **Tinnitus cause : head trauma** | | | | Chi2 = 0.1 | 1.0 | 0.004 (Negligible) | | | |
| Yes | 535 (97.6%) | 2165 (97.6%) | 282 (97.9%) | | | | | | |
| No | 13 (2.4%) | 53 (2.4%) | 6 (2.1%) | | | | | | |
| Missing | 94 (14.6%) | 433 (16.3%) | 46 (13.8%) | | | | | | |
| **Tinnitus cause : noise trauma** | | | | Chi2 = 3.6 | 1.0 | 0.022 (Negligible) | | | |
| Yes | 464 (84.7%) | 1830 (82.5%) | 229 (79.5%) | | | | | | |
| No | 84 (15.3%) | 388 (17.5%) | 59 (20.5%) | | | | | | |
| Missing | 94 (14.6%) | 433 (16.3%) | 46 (13.8%) | | | | | | |
| **Tinnitus cause : other** | | | | Chi2 = 4.0 | 1.0 | 0.024 (Negligible) | | | |
| Yes | 280 (51.1%) | 1178 (53.1%) | 168 (58.3%) | | | | | | |
| No | 268 (48.9%) | 1040 (46.9%) | 120 (41.7%) | | | | | | |
| Missing | 94 (14.6%) | 433 (16.3%) | 46 (13.8%) | | | | | | |
| **Tinnitus varies from day to day** | | | | Chi2 = 136.9** | $p < 0.001$ | 0.137 (Small) | | | |
| Yes | 489 (77.1%) | 1485 (56.8%) | 264 (80%) | | | | TW VS NE | $p < 0.001$ | 0.155 (Small) |
| No | 145 (22.9%) | 1128 (43.2%) | 66 (20%) | | | | TI VS NE | $p < 0.001$ | 0.133 (Small) |
| Missing | 8 (1.2%) | 38 (1.4%) | 4 (1.2%) | | | | TW VS TI | 1.0 | 0.016 (Negligible) |
| **Tinnitus intermittent or continuous ?** | | | | Chi2 = 38.9** | $p < 0.001$ | 0.073 (Small) | | | |
| Intermittent | 112 (17.8%) | 307 (11.7%) | 74 (22.4%) | | | | TW VS NE | 0.002 | 0.067 (Negligible) |
| Continuous | 517 (82.2%) | 2317 (88.3%) | 256 (77.6%) | | | | TI VS NE | $p < 0.001$ | 0.09 (Negligible) |
| Missing | 13 (2%) | 27 (1%) | 4 (1.2%) | | | | TW VS TI | 1.0 | 0.027 (Negligible) |
| **Influence of stress over tinnitus** | | | | Chi2 = 84.2** | $p < 0.001$ | 0.076 (Small) | | | |
| Worsens | 477 (77.2%) | 1742 (67.5%) | 294 (89.6%) | | | | TW VS NE | $p < 0.001$ | 0.057 (Negligible) |
| Improves | 8 (1.3%) | 29 (1.1%) | 0 (0%) | | | | TI VS NE | $p < 0.001$ | 0.097 (Small) |
| No effect | 133 (21.5%) | 809 (31.4%) | 34 (10.4%) | | | | TW VS TI | $p < 0.001$ | 0.057 (Negligible) |
| Missing | 24 (3.7%) | 71 (2.7%) | 6 (1.8%) | | | | | | |
| **Sleep at night and tinnitus during the day** | | | | Chi2 = 399.0** | $p < 0.001$ | 0.235 (Medium) | | | |

| | | | | | | | | |
|---|---|---|---|---|---|---|---|---|
| Linked | 261 (71.5%) | 338 (22.8%) | 127 (68.6%) | | | TW VS NE | p < 0.001 | 0.295 (Small) |
| Not linked | 104 (28.5%) | 1145 (77.2%) | 58 (31.4%) | | | TI VS NE | p < 0.001 | 0.216 (Small) |
| Missing | 277 (43.1%) | 1168 (44.1%) | 149 (44.6%) | | | TW VS TI | 1.0 | 0.01 (Negligible) |
| **Some sounds can worsen tinnitus** | | | | Chi2 = 34.4** | p < 0.001 | 0.069 (Negligible) | | |
| Yes | 374 (73.5%) | 1407 (66%) | 242 (81.2%) | | | TW VS NE | 0.048 | 0.053 (Negligible) |
| No | 135 (26.5%) | 725 (34%) | 56 (18.8%) | | | TI VS NE | p < 0.001 | 0.086 (Negligible) |
| Missing | 133 (20.7%) | 519 (19.6%) | 36 (10.8%) | | | TW VS TI | 0.385 | 0.04 (Negligible) |
| **Some sounds can suppress tinnitus** | | | | Chi2 = 15.5** | 0.015 | 0.046 (Negligible) | | |
| Yes | 431 (78.9%) | 1665 (74.7%) | 238 (84.4%) | | | TW VS NE | 0.859 | 0.033 (Negligible) |
| No | 115 (21.1%) | 564 (25.3%) | 44 (15.6%) | | | TI VS NE | 0.016 | 0.058 (Negligible) |
| Missing | 96 (15%) | 422 (15.9%) | 52 (15.6%) | | | TW VS TI | 1.0 | 0.03 (Negligible) |
| **Some sounds can cause physical discomfort** | | | | Chi2 = 14.3** | 0.025 | 0.044 (Negligible) | | |
| Yes | 359 (62.7%) | 1357 (58.9%) | 208 (69.8%) | | | TW VS NE | 1.0 | 0.026 (Negligible) |
| No | 214 (37.3%) | 946 (41.1%) | 90 (30.2%) | | | TI VS NE | 0.014 | 0.059 (Negligible) |
| Missing | 69 (10.7%) | 348 (13.1%) | 36 (10.8%) | | | TW VS TI | 0.859 | 0.034 (Negligible) |
| **Somatosensory : Jaw or head movements can modulate tinnitus** | | | | Chi2 = 98.6** | p < 0.001 | 0.117 (Small) | | |
| Yes | 325 (51.5%) | 835 (32%) | 153 (46.6%) | | | TW VS NE | p < 0.001 | 0.152 (Small) |
| No | 306 (48.5%) | 1777 (68%) | 175 (53.4%) | | | TI VS NE | p < 0.001 | 0.087 (Negligible) |
| Missing | 11 (1.7%) | 39 (1.5%) | 6 (1.8%) | | | TW VS TI | 1.0 | 0.023 (Negligible) |
| **Temporomandibular disorder** | | | | Chi2 = 12.1* | 0.058 | 0.041 (Negligible) | | |
| Yes | 176 (28%) | 646 (24.7%) | 109 (33.1%) | | | | | |
| No | 453 (72%) | 1966 (75.3%) | 220 (66.9%) | | | | | |
| Missing | 13 (2%) | 39 (1.5%) | 5 (1.5%) | | | | | |
| **Neck pain** | | | | Chi2 = 0.6 | 1.0 | 0.009 (Negligible) | | |
| Yes | 364 (58.1%) | 1500 (58%) | 198 (60.2%) | | | | | |
| No | 263 (41.9%) | 1088 (42%) | 131 (39.8%) | | | | | |
| Missing | 15 (2.3%) | 63 (2.4%) | 5 (1.5%) | | | | | |
| **Headaches** | | | | Chi2 = 3.7 | 1.0 | 0.023 (Negligible) | | |
| Yes | 256 (40.8%) | 993 (38.4%) | 142 (43.3%) | | | | | |
| No | 372 (59.2%) | 1596 (61.6%) | 186 (56.7%) | | | | | |
| Missing | 14 (2.2%) | 62 (2.3%) | 6 (1.8%) | | | | | |
| **Vertigo** | | | | Chi2 = 9.5* | 0.154 | 0.036 (Negligible) | | |
| Yes | 200 (32.3%) | 876 (34.1%) | 136 (42%) | | | | | |
| No | 419 (67.7%) | 1690 (65.9%) | 188 (58%) | | | | | |
| Missing | 23 (3.6%) | 85 (3.2%) | 10 (3%) | | | | | |
| **Number of treatment tested** | | | | Chi2 = 36.0** | 0.003 | 0.032 (Negligible) | | |
| 0 (none) | 76 (12%) | 392 (15%) | 63 (19%) | | | TW VS NE | 0.002 | 0.039 (Negligible) |
| One | 83 (13.1%) | 441 (16.9%) | 48 (14.5%) | | | TI VS NE | 1.0 | 0.021 (Negligible) |
| 2 to 4 | 175 (27.7%) | 776 (29.7%) | 92 (27.8%) | | | TW VS TI | 0.175 | 0.03 (Negligible) |
| 5 and more | 148 (23.4%) | 453 (17.3%) | 67 (20.2%) | | | | | |
| several | 42 (6.6%) | 102 (3.9%) | 8 (2.4%) | | | | | |
| many | 108 (17.1%) | 453 (17.3%) | 53 (16%) | | | | | |

| | | | | | | | | | |
|---|---|---|---|---|---|---|---|---|---|
| Missing | 10 (1.6%) | 34 (1.3%) | 3 (0.9%) | | | | | | |
| **Currently under psychiatric treatment** | | | | Chi2 = 2.8 | 1.0 | 0.02 (Negligible) | | | |
| Yes | 135 (21.3%) | 481 (18.4%) | 63 (19%) | | | | | | |
| No | 500 (78.7%) | 2139 (81.6%) | 268 (81%) | | | | | | |
| Missing | 7 (1.1%) | 31 (1.2%) | 3 (0.9%) | | | | | | |
| **Hearing aid user** | | | | Chi2 = 24.9** | 0.013 | 0.034 (Negligible) | | | |
| right | 15 (2.4%) | 42 (1.6%) | 13 (4%) | | | | TW VS NE | 0.036 | 0.039 (Negligible) |
| left | 20 (3.2%) | 68 (2.6%) | 11 (3.4%) | | | | TI VS NE | 0.417 | 0.03 (Negligible) |
| both | 98 (15.6%) | 274 (10.5%) | 33 (10.3%) | | | | TW VS TI | 1.0 | 0.025 (Negligible) |
| none | 495 (78.8%) | 2220 (85.3%) | 264 (82.2%) | | | | | | |
| Missing | 14 (2.2%) | 47 (1.8%) | 13 (3.9%) | | | | | | |

**Supplementary table 2**. Extended group comparison in the Tinnitus Hub survey database. *SD : Standard Deviation, TW : Tinnitus worsens after naps, NE : No effect of naps on tinnitus, TI : Tinnitus improves after naps, TMJ : temporo-mandibular joint, \* : p < 0.05 before Holm correction, \*\* : p < 0.05 after Holm correction.*

| | Worsens N= 1404 | No effect N= 4204 | Improves N= 507 | Statistic | p-Value | Effect size | Post-hoc test | | Post-hoc effect size |
|---|---|---|---|---|---|---|---|---|---|
| **Age (years)** | | | | H = 17.6** | 0.009 | 0.003 (Negligible) | | | |
| Mean (SD) | 54 (148) | 54.4 (13.6) | 51.7 (14.54) | | | | TW VS NE | 1.0 | 0.03 (Negligible) |
| Median [Min Max] | 57 [14, 93] | 57 [6, 96] | 54 [15, 94] | | | | TI VS NE | 0.002 | 0.197 (Negligible) |
| Missing | 0 (0%) | 0 (0%) | 0 (0%) | | | | TW VS TI | 0.039 | 0.161 (Negligible) |
| **Tinnitus duration (months)** | | | | H = 16.2** | 0.016 | 0.002 (Negligible) | | | |
| Mean (SD) | 93.8 (100.93) | 106 (113.33) | 89 (105.63) | | | | TW VS NE | 1.0 | 0.11 (Negligible) |
| Median [Min Max] | 52 [1, 371] | 59 [0, 371] | 42 [1, 371] | | | | TI VS NE | 0.007 | 0.151 (Negligible) |
| Missing | 2 (0.1%) | 2 (0%) | 0 (0%) | | | | TW VS TI | 0.796 | 0.047 (Negligible) |
| **Average loudness (0-10)** | | | | H = 8.9* | 0.471 | 0.001 (Negligible) | | | |
| Mean (SD) | 5.8 (2.33) | 5.6 (2.35) | 5.6 (2.13) | | | | | | |
| Median [Min Max] | 6 [1, 10] | 6 [0, 10] | 6 [0, 10] | | | | | | |
| Missing | 7 (0.5%) | 17 (0.4%) | 0 (0%) | | | | | | |
| **Loudness on good days (0-10)** | | | | H = 9.3* | 0.404 | 0.001 (Negligible) | | | |
| Mean (SD) | 3.6 (2.14) | 3.7 (2.24) | 3.4 (1.9) | | | | | | |
| Median [Min Max] | 3 [0, 10] | 3 [0, 10] | 3 [0, 10] | | | | | | |
| Missing | 13 (0.9%) | 39 (0.9%) | 3 (0.6%) | | | | | | |
| **Loudness on bad days (0-10)** | | | | H = 8.6* | 0.525 | 0.001 (Negligible) | | | |
| Mean (SD) | 6.6 (33) | 6.4 (2.87) | 6.6 (2.81) | | | | | | |
| Median [Min Max] | 7 [0, 10] | 7 [0, 10] | 7 [1, 10] | | | | | | |
| Missing | 30 (2.1%) | 73 (1.7%) | 7 (1.4%) | | | | | | |
| **Current loudness (0-10)** | | | | H = 1.6 | 1.0 | -0.0 (Negligible) | | | |
| Mean (SD) | 5.3 (2.51) | 5.2 (2.54) | 5.1 (2.34) | | | | | | |
| Median [Min Max] | 5 [0, 10] | 5 [0, 10] | 5 [0, 10] | | | | | | |
| Missing | 38 (2.7%) | 81 (1.9%) | 13 (2.6%) | | | | | | |
| **Average annoyance (0-10)** | | | | H = 30.5** | p < 0.001 | 0.005 (Negligible) | | | |
| Mean (SD) | 5.9 (2.99) | 5.4 (36) | 5.6 (2.94) | | | | TW VS NE | p < 0.001 | 0.169 (Negligible) |
| Median [Min Max] | 6 [0, 10] | 5 [0, 10] | 5 [0, 10] | | | | TI VS NE | 1.0 | 0.054 (Negligible) |
| Missing | 8 (0.6%) | 12 (0.3%) | 2 (0.4%) | | | | TW VS TI | 1.0 | 0.118 (Negligible) |
| **Annoyance on good days (0-10)** | | | | H = 3.3 | 1.0 | 0.0 (Negligible) | | | |
| Mean (SD) | 3.6 (2.89) | 3.5 (2.95) | 3.3 (2.62) | | | | | | |
| Median [Min Max] | 3 [0, 10] | 3 [0, 10] | 3 [0, 10] | | | | | | |
| Missing | 20 (1.4%) | 36 (0.9%) | 4 (0.8%) | | | | | | |
| **Annoyance on bad days (0-10)** | | | | H = 24.5** | p < 0.001 | 0.004 (Negligible) | | | |
| Mean (SD) | 7 (3.2) | 6.5 (3.2) | 6.6 (3.19) | | | | TW VS NE | p < 0.001 | 0.138 (Negligible) |
| Median [Min Max] | 8 [0, 10] | 7 [0, 10] | 7 [0, 10] | | | | TI VS NE | 1.0 | 0.035 (Negligible) |
| Missing | 21 (1.5%) | 69 (1.6%) | 4 (0.8%) | | | | TW VS TI | 0.914 | 0.103 (Negligible) |
| **Current annoyance (0-10)** | | | | H = 7.2* | 0.952 | 0.001 (Negligible) | | | |
| Mean (SD) | 5.3 (34) | 5.1 (39) | 5 (2.89) | | | | | | |
| Median [Min Max] | 5 [0, 10] | 5 [0, 10] | 5 [0, 10] | | | | | | |
| Missing | 27 (1.9%) | 76 (1.8%) | 11 (2.2%) | | | | | | |
| **% of time aware of tinnitus (0-100)** | | | | H = 26.6** | p < 0.001 | 0.004 (Negligible) | | | |
| Mean (SD) | 71.9 (26.17) | 67.7 (28.15) | 65.9 (26.81) | | | | TW VS NE | p < 0.001 | 0.15 (Negligible) |
| Median [Min Max] | 76 [0, 100] | 75 [0, 100] | 70 [0, 100] | | | | TI VS NE | 1.0 | 0.065 (Negligible) |
| Missing | 0 (0%) | 0 (0%) | 0 (0%) | | | | TW VS TI | p < 0.001 | 0.226 (Small) |
| **Hearing loss grade (0 : none, 3 : severe)** | | | | H = 3.1 | 1.0 | 0.0 (Negligible) | | | |
| Mean (SD) | 1.2 (0.81) | 1.1 (0.8) | 1.1 (0.85) | | | | | | |
| Median [Min Max] | 1 [0, 3] | 1 [0, 3] | 1 [0, 3] | | | | | | |
| Missing | 0 (0%) | 0 (0%) | 0 (0%) | | | | | | |
| **Frequency of fleeting tinnitus (0 : never, 5 : daily)** | | | | H = 60.0** | p < 0.001 | 0.009 (Negligible) | | | |
| Mean (SD) | 2.9 (1.82) | 2.5 (1.91) | 2.8 (1.89) | | | | TW VS NE | p < 0.001 | 0.228 (Small) |
| Median [Min Max] | 4 [0, 5] | 3 [0, 5] | 3 [0, 5] | | | | TI VS NE | 0.022 | 0.161 (Negligible) |
| Missing | 0 (0%) | 0 (0%) | 0 (0%) | | | | TW VS TI | 1.0 | 0.067 (Negligible) |
| **Gender** | | | | Chi2 = 59.6** | p < 0.001 | 0.07 (Small) | | | |
| Male | 832 (59.7%) | 2031 (48.6%) | 291 (57.9%) | | | | TW VS NE | p < 0.001 | 0.092 (Negligible) |
| Female | 562 (40.3%) | 2151 (51.4%) | 212 (42.1%) | | | | TI VS NE | 0.005 | 0.05 (Negligible) |
| Missing | 10 (0.7%) | 22 (0.5%) | 4 (0.8%) | | | | TW VS TI | 1.0 | 0.008 (Negligible) |
| **Tinnitus sound : A pure tone** | | | | Chi2 = 12.1* | 0.113 | 0.031 (Negligible) | | | |
| No | 955 (68%) | 2651 (63.1%) | 336 (66.3%) | | | | | | |
| Yes | 449 (32%) | 1553 (36.9%) | 171 (33.7%) | | | | | | |
| Missing | 0 (0%) | 0 (0%) | 0 (0%) | | | | | | |
| **Tinnitus sound : A mixture of tones** | | | | Chi2 = 21.7** | 0.001 | 0.042 (Negligible) | | | |
| No | 904 (64.4%) | 2977 (70.8%) | 338 (66.7%) | | | | TW VS NE | p < 0.001 | 0.057 (Negligible) |
| Yes | 500 (35.6%) | 1227 (29.2%) | 169 (33.3%) | | | | TI VS NE | 1.0 | 0.024 (Negligible) |
| Missing | 0 (0%) | 0 (0%) | 0 (0%) | | | | TW VS TI | 1.0 | 0.011 (Negligible) |
| **Tinnitus sound : A low buzzing** | | | | Chi2 = 2.4 | 1.0 | 0.014 (Negligible) | | | |
| No | 1213 (86.4%) | 3673 (87.4%) | 432 (85.2%) | | | | | | |
| Yes | 191 (13.6%) | 531 (12.6%) | 75 (14.8%) | | | | | | |
| Missing | 0 (0%) | 0 (0%) | 0 (0%) | | | | | | |
| **Tinnitus sound : A high buzzing** | | | | Chi2 = 8.1* | 0.659 | 0.026 (Negligible) | | | |
| No | 905 (64.5%) | 2748 (65.4%) | 299 (59%) | | | | | | |
| Yes | 499 (35.5%) | 1456 (34.6%) | 208 (41%) | | | | | | |
| Missing | 0 (0%) | 0 (0%) | 0 (0%) | | | | | | |
| **Tinnitus sound : electric/interference** | | | | Chi2 = 17.3** | 0.01 | 0.038 (Negligible) | | | |
| No | 1041 (74.1%) | 3337 (79.4%) | 402 (79.3%) | | | | TW VS NE | 0.003 | 0.052 (Negligible) |
| Yes | 363 (25.9%) | 867 (20.6%) | 105 (20.7%) | | | | TI VS NE | 1.0 | 0.0 (Negligible) |
| Missing | 0 (0%) | 0 (0%) | 0 (0%) | | | | TW VS TI | 0.932 | 0.029 (Negligible) |
| **Tinnitus sound : A low rumbling** | | | | Chi2 = 3.3 | 1.0 | 0.016 (Negligible) | | | |
| No | 1325 (94.4%) | 4014 (95.5%) | 486 (95.9%) | | | | | | |
| Yes | 79 (5.6%) | 190 (4.5%) | 21 (4.1%) | | | | | | |
| Missing | 0 (0%) | 0 (0%) | 0 (0%) | | | | | | |

| | | | | | | |
|---|---|---|---|---|---|---|
| **Tinnitus sound : A static noise** | | | | Chi2 = 9.7* | 0.34 | 0.028 (Negligible) |
| No | 1124 (80.1%) | 3492 (83.1%) | 401 (79.1%) | | | |
| Yes | 280 (19.9%) | 712 (16.9%) | 106 (20.9%) | | | |
| Missing | 0 (0%) | 0 (0%) | 0 (0%) | | | |
| **Tinnitus sound : clicking** | | | | Chi2 = 7.4* | 0.882 | 0.025 (Negligible) |
| No | 1311 (93.4%) | 3983 (94.7%) | 468 (92.3%) | | | |
| Yes | 93 (6.6%) | 221 (5.3%) | 39 (7.7%) | | | |
| Missing | 0 (0%) | 0 (0%) | 0 (0%) | | | |
| **Tinnitus sound : beeping (morse code)** | | | | Chi2 = 1.6 | 1.0 | 0.011 (Negligible) |
| No | 1335 (95.1%) | 4027 (95.8%) | 482 (95.1%) | | | |
| Yes | 69 (4.9%) | 177 (4.2%) | 25 (4.9%) | | | |
| Missing | 0 (0%) | 0 (0%) | 0 (0%) | | | |
| **Tinnitus sound : A pulsatile whooshing noise** | | | | Chi2 = 2.0 | 1.0 | 0.013 (Negligible) |
| No | 1250 (89%) | 3786 (90.1%) | 449 (88.6%) | | | |
| Yes | 154 (11%) | 418 (9.9%) | 58 (11.4%) | | | |
| Missing | 0 (0%) | 0 (0%) | 0 (0%) | | | |
| **Tinnitus sound : A non-pulsatile whooshing noise** | | | | Chi2 = 6.1* | 1.0 | 0.022 (Negligible) |
| No | 1285 (91.5%) | 3915 (93.1%) | 461 (90.9%) | | | |
| Yes | 119 (8.5%) | 289 (6.9%) | 46 (9.1%) | | | |
| Missing | 0 (0%) | 0 (0%) | 0 (0%) | | | |
| **Tinnitus sound : pulsatile sound** | | | | Chi2 = 2.0 | 1.0 | 0.013 (Negligible) |
| No | 1287 (91.7%) | 3801 (90.4%) | 461 (90.9%) | | | |
| Yes | 117 (8.3%) | 403 (9.6%) | 46 (9.1%) | | | |
| Missing | 0 (0%) | 0 (0%) | 0 (0%) | | | |
| **Tinnitus sound : other** | | | | Chi2 = 3.8 | 1.0 | 0.018 (Negligible) |
| No | 1170 (83.3%) | 3499 (83.2%) | 439 (86.6%) | | | |
| Yes | 234 (16.7%) | 705 (16.8%) | 68 (13.4%) | | | |
| Missing | 0 (0%) | 0 (0%) | 0 (0%) | | | |
| **Tinnitus cause : noise trauma** | | | | Chi2 = 1.5 | 1.0 | 0.011 (Negligible) |
| No | 1169 (83.3%) | 3543 (84.3%) | 419 (82.6%) | | | |
| Yes | 235 (16.7%) | 661 (15.7%) | 88 (17.4%) | | | |
| Missing | 0 (0%) | 0 (0%) | 0 (0%) | | | |
| **Tinnitus cause : hearing loss** | | | | Chi2 = 4.8 | 1.0 | 0.02 (Negligible) |
| No | 1106 (78.8%) | 3393 (80.7%) | 421 (83%) | | | |
| Yes | 298 (21.2%) | 811 (19.3%) | 86 (17%) | | | |
| Missing | 0 (0%) | 0 (0%) | 0 (0%) | | | |
| **Tinnitus cause : age-related hearing loss** | | | | Chi2 = 6.4* | 1.0 | 0.023 (Negligible) |
| No | 1228 (87.5%) | 3741 (89%) | 434 (85.6%) | | | |
| Yes | 176 (12.5%) | 463 (11%) | 73 (14.4%) | | | |
| Missing | 0 (0%) | 0 (0%) | 0 (0%) | | | |
| **Tinnitus cause : sudden hearing loss** | | | | Chi2 = 6.2* | 1.0 | 0.023 (Negligible) |
| No | 1321 (94.1%) | 3946 (93.9%) | 462 (91.1%) | | | |
| Yes | 83 (5.9%) | 258 (6.1%) | 45 (8.9%) | | | |

| | | | | | | | | | |
|---|---|---|---|---|---|---|---|---|---|
| Missing | 0 (0%) | 0 (0%) | 0 (0%) | | | | | | |
| **Tinnitus cause : Meniere disease** | | | | Chi2 = 6.6* | 1.0 | 0.023 (Negligible) | | | |
| No | 1344 (95.7%) | 4016 (95.5%) | 472 (93.1%) | | | | | | |
| Yes | 60 (4.3%) | 188 (4.5%) | 35 (6.9%) | | | | | | |
| Missing | 0 (0%) | 0 (0%) | 0 (0%) | | | | | | |
| **Tinnitus cause : head or neck injury** | | | | Chi2 = 26.0** | p < 0.001 | 0.046 (Negligible) | | | |
| No | 1228 (87.5%) | 3859 (91.8%) | 448 (88.4%) | | | | TW VS NE | p < 0.001 | 0.061 (Negligible) |
| Yes | 176 (12.5%) | 345 (8.2%) | 59 (11.6%) | | | | TI VS NE | 0.525 | 0.032 (Negligible) |
| Missing | 0 (0%) | 0 (0%) | 0 (0%) | | | | TW VS TI | 1.0 | 0.006 (Negligible) |
| **Tinnitus cause : barotrauma** | | | | Chi2 = 5.1 | 1.0 | 0.02 (Negligible) | | | |
| No | 1374 (97.9%) | 4073 (96.9%) | 497 (98%) | | | | | | |
| Yes | 30 (2.1%) | 131 (3.1%) | 10 (2%) | | | | | | |
| Missing | 0 (0%) | 0 (0%) | 0 (0%) | | | | | | |
| **Tinnitus cause : TMJ dysfunction** | | | | Chi2 = 9.9* | 0.317 | 0.028 (Negligible) | | | |
| No | 1242 (88.5%) | 3814 (90.7%) | 443 (87.4%) | | | | | | |
| Yes | 162 (11.5%) | 390 (9.3%) | 64 (12.6%) | | | | | | |
| Missing | 0 (0%) | 0 (0%) | 0 (0%) | | | | | | |
| **Tinnitus cause : psychological (stress, anxiety, depression)** | | | | Chi2 = 60.1** | p < 0.001 | 0.07 (Small) | | | |
| No | 1101 (78.4%) | 3593 (85.5%) | 382 (75.3%) | | | | TW VS NE | p < 0.001 | 0.079 (Negligible) |
| Yes | 303 (21.6%) | 611 (14.5%) | 125 (24.7%) | | | | TI VS NE | p < 0.001 | 0.075 (Negligible) |
| Missing | 0 (0%) | 0 (0%) | 0 (0%) | | | | TW VS TI | 1.0 | 0.017 (Negligible) |
| **Tinnitus cause : ototoxicity** | | | | Chi2 = 18.1** | 0.007 | 0.038 (Negligible) | | | |
| No | 1212 (86.3%) | 3799 (90.4%) | 451 (89%) | | | | TW VS NE | 0.001 | 0.054 (Negligible) |
| Yes | 192 (13.7%) | 405 (9.6%) | 56 (11%) | | | | TI VS NE | 1.0 | 0.012 (Negligible) |
| Missing | 0 (0%) | 0 (0%) | 0 (0%) | | | | TW VS TI | 1.0 | 0.018 (Negligible) |
| **Tinnitus cause : otosclerosis** | | | | Chi2 = 0.4 | 1.0 | 0.006 (Negligible) | | | |
| No | 1387 (98.8%) | 4150 (98.7%) | 499 (98.4%) | | | | | | |
| Yes | 17 (1.2%) | 54 (1.3%) | 8 (1.6%) | | | | | | |
| Missing | 0 (0%) | 0 (0%) | 0 (0%) | | | | | | |
| **Tinnitus cause : eustachian tube dysfunction** | | | | Chi2 = 1.0 | 1.0 | 0.009 (Negligible) | | | |
| No | 1312 (93.4%) | 3947 (93.9%) | 471 (92.9%) | | | | | | |
| Yes | 92 (6.6%) | 257 (6.1%) | 36 (7.1%) | | | | | | |
| Missing | 0 (0%) | 0 (0%) | 0 (0%) | | | | | | |
| **Tinnitus cause : dental treatment** | | | | Chi2 = 2.2 | 1.0 | 0.013 (Negligible) | | | |
| No | 1368 (97.4%) | 4069 (96.8%) | 488 (96.3%) | | | | | | |
| Yes | 36 (2.6%) | 135 (3.2%) | 19 (3.7%) | | | | | | |
| Missing | 0 (0%) | 0 (0%) | 0 (0%) | | | | | | |
| **Tinnitus cause : allergy** | | | | Chi2 = 3.0 | 1.0 | 0.016 (Negligible) | | | |
| No | 1365 (97.2%) | 4049 (96.3%) | 492 (97%) | | | | | | |
| Yes | 39 (2.8%) | 155 (3.7%) | 15 (3%) | | | | | | |
| Missing | 0 (0%) | 0 (0%) | 0 (0%) | | | | | | |
| **Tinnitus cause : ear wax procedure (syringing, candling...)** | | | | Chi2 = 0.8 | 1.0 | 0.008 (Negligible) | | | |
| No | 1366 (97.3%) | 4106 (97.7%) | 496 (97.8%) | | | | | | |
| Yes | 38 (2.7%) | 98 (2.3%) | 11 (2.2%) | | | | | | |
| Missing | 0 (0%) | 0 (0%) | 0 (0%) | | | | | | |
| **Tinnitus cause : metabolic (diabetes, thyroid, B12, hyperlipidaemia etc.)** | | | | Chi2 = 13.4* | 0.064 | 0.033 (Negligible) | | | |
| No | 1368 (97.4%) | 4069 (96.8%) | 477 (94.1%) | | | | | | |
| Yes | 36 (2.6%) | 135 (3.2%) | 30 (5.9%) | | | | | | |
| Missing | 0 (0%) | 0 (0%) | 0 (0%) | | | | | | |
| **Tinnitus cause : virus or infection** | | | | Chi2 = 6.0 | 1.0 | 0.022 (Negligible) | | | |
| No | 1186 (84.5%) | 3618 (86.1%) | 418 (82.4%) | | | | | | |
| Yes | 218 (15.5%) | 586 (13.9%) | 89 (17.6%) | | | | | | |
| Missing | 0 (0%) | 0 (0%) | 0 (0%) | | | | | | |
| **Tinnitus cause : ear wax build up** | | | | Chi2 = 0.6 | 1.0 | 0.007 (Negligible) | | | |
| No | 1366 (97.3%) | 4086 (97.2%) | 490 (96.6%) | | | | | | |
| Yes | 38 (2.7%) | 118 (2.8%) | 17 (3.4%) | | | | | | |
| Missing | 0 (0%) | 0 (0%) | 0 (0%) | | | | | | |
| **Tinnitus cause : unknown** | | | | Chi2 = 18.8** | 0.005 | 0.039 (Negligible) | | | |
| No | 72 (5.1%) | 122 (2.9%) | 26 (5.1%) | | | | TW VS NE | 0.006 | 0.049 (Negligible) |
| Yes | 1332 (94.9%) | 4082 (97.1%) | 481 (94.9%) | | | | TI VS NE | 0.455 | 0.033 (Negligible) |
| Missing | 0 (0%) | 0 (0%) | 0 (0%) | | | | TW VS TI | 1.0 | 0.0 (Negligible) |
| **Fluctuations of tinnitus** | | | | Chi2 = 257.3** | p < 0.001 | 0.084 (Small) | | | |
| No fluctuations | 268 (19.1%) | 1512 (36%) | 80 (15.8%) | | | | TW VS NE | p < 0.001 | 0.092 (Small) |
| Grows louder as day progresses | 703 (50.1%) | 1782 (42.4%) | 218 (43%) | | | | TI VS NE | p < 0.001 | 0.084 (Small) |
| Grows quieter as day progresses | 122 (8.7%) | 193 (4.6%) | 36 (7.1%) | | | | TW VS TI | p < 0.001 | 0.039 (Negligible) |
| Changes within the day or over days without any pattern | 311 (22.2%) | 717 (17.1%) | 173 (34.1%) | | | | | | |
| Missing | 0 (0%) | 0 (0%) | 0 (0%) | | | | | | |
| **Influence of stress over tinnitus** | | | | Chi2 = 292.6** | p < 0.001 | 0.109 (Small) | | | |
| Worsens | 1128 (80.3%) | 2439 (58%) | 409 (80.7%) | | | | TW VS NE | p < 0.001 | 0.136 (Small) |
| No effect | 271 (19.3%) | 1746 (41.5%) | 95 (18.7%) | | | | TI VS NE | p < 0.001 | 0.09 (Small) |
| Improves | 5 (0.4%) | 19 (0.5%) | 3 (0.6%) | | | | TW VS TI | 1.0 | 0.007 (Negligible) |
| Missing | 0 (0%) | 0 (0%) | 0 (0%) | | | | | | |
| **Influence of anxiety over tinnitus** | | | | Chi2 = 317.5** | p < 0.001 | 0.114 (Small) | | | |
| Worsens | 1090 (77.6%) | 2272 (54%) | 394 (77.7%) | | | | TW VS NE | p < 0.001 | 0.141 (Small) |
| No effect | 309 (22%) | 1910 (45.4%) | 107 (21.1%) | | | | TI VS NE | p < 0.001 | 0.095 (Small) |
| Improves | 5 (0.4%) | 22 (0.5%) | 6 (1.2%) | | | | TW VS TI | 1.0 | 0.019 (Negligible) |
| Missing | 0 (0%) | 0 (0%) | 0 (0%) | | | | | | |
| **Influence of a good night sleep over tinnitus** | | | | Chi2 = 1373.5** | p < 0.001 | 0.237 (Medium) | | | |
| Worsens | 476 (33.9%) | 272 (6.5%) | 24 (4.7%) | | | | TW VS NE | p < 0.001 | 0.257 (Medium) |
| No effect | 430 (30.6%) | 2665 (63.4%) | 67 (13.2%) | | | | TI VS NE | p < 0.001 | 0.21 (Medium) |
| Improves | 498 (35.5%) | 1267 (30.1%) | 416 (82.1%) | | | | TW VS TI | p < 0.001 | 0.165 (Small) |
| Missing | 0 (0%) | 0 (0%) | 0 (0%) | | | | | | |
| **Influence of poor sleep over tinnitus** | | | | Chi2 = 694.5** | p < 0.001 | 0.169 (Medium) | | | |
| Worsens | 1134 (80.8%) | 1844 (43.9%) | 379 (74.8%) | | | | TW VS NE | p < 0.001 | 0.22 (Medium) |

| | | | | | | | | |
|---|---|---|---|---|---|---|---|---|
| No effect | 250 (17.8%) | 2320 (55.2%) | 118 (23.3%) | | | TI VS NE | p < 0.001 | 0.123 (Small) |
| Improves | 20 (1.4%) | 40 (1%) | 10 (2%) | | | TW VS TI | 0.706 | 0.026 (Negligible) |
| Missing | 0 (0%) | 0 (0%) | 0 (0%) | | | | | |
| **Some sounds can worsen tinnitus** | | | | Chi2 = 20.0** | 0.003 | 0.04 (Negligible) | | |
| No | 663 (47.2%) | 2268 (53.9%) | 254 (50.1%) | | | TW VS NE | p < 0.001 | 0.055 (Negligible) |
| Yes | 741 (52.8%) | 1936 (46.1%) | 253 (49.9%) | | | TI VS NE | 1.0 | 0.02 (Negligible) |
| Missing | 0 (0%) | 0 (0%) | 0 (0%) | | | TW VS TI | 1.0 | 0.014 (Negligible) |
| **Some sounds can reduce tinnitus** | | | | Chi2 = 36.6** | p < 0.001 | 0.055 (Negligible) | | |
| No | 426 (30.3%) | 1002 (23.8%) | 167 (32.9%) | | | TW VS NE | p < 0.001 | 0.062 (Negligible) |
| Yes | 978 (69.7%) | 3202 (76.2%) | 340 (67.1%) | | | TI VS NE | p < 0.001 | 0.057 (Negligible) |
| Missing | 0 (0%) | 0 (0%) | 0 (0%) | | | TW VS TI | 1.0 | 0.013 (Negligible) |
| **Tinnitus sound masking** | | | | Chi2 = 128.7** | p < 0.001 | 0.046 (Small) | | |
| No masking | 101 (7.2%) | 410 (9.8%) | 69 (13.6%) | | | TW VS NE | p < 0.001 | 0.058 (Small) |
| Only a small selection of specific sounds | 162 (11.5%) | 357 (8.5%) | 63 (12.4%) | | | TI VS NE | p < 0.001 | 0.033 (Negligible) |
| Shower / water noises | 287 (20.4%) | 485 (11.5%) | 68 (13.4%) | | | TW VS TI | p < 0.001 | 0.033 (Negligible) |
| TV, Music or general background noise | 334 (23.8%) | 1291 (30.7%) | 135 (26.6%) | | | | | |
| White noise or special masking noises | 197 (14%) | 555 (13.2%) | 80 (15.8%) | | | | | |
| Masked by nearly all sounds | 323 (23%) | 1106 (26.3%) | 92 (18.1%) | | | | | |
| Missing | 0 (0%) | 0 (0%) | 0 (0%) | | | | | |
| **Influence of intense workout over tinnitus** | | | | Chi2 = 252.2** | p < 0.001 | 0.102 (Small) | | |
| Worsens | 424 (30.2%) | 645 (15.3%) | 126 (24.9%) | | | TW VS NE | p < 0.001 | 0.121 (Small) |
| No effect | 847 (60.3%) | 3282 (78.1%) | 295 (58.2%) | | | TI VS NE | p < 0.001 | 0.096 (Small) |
| Improves | 133 (9.5%) | 277 (6.6%) | 86 (17%) | | | TW VS TI | p < 0.001 | 0.043 (Negligible) |
| Missing | 0 (0%) | 0 (0%) | 0 (0%) | | | | | |
| **Influence of moderate workout over tinnitus** | | | | Chi2 = 326.7** | p < 0.001 | 0.116 (Small) | | |
| Worsens | 423 (30.1%) | 584 (13.9%) | 117 (23.1%) | | | TW VS NE | p < 0.001 | 0.135 (Small) |
| No effect | 801 (57.1%) | 3219 (76.6%) | 267 (52.7%) | | | TI VS NE | p < 0.001 | 0.11 (Small) |
| Improves | 180 (12.8%) | 401 (9.5%) | 123 (24.3%) | | | TW VS TI | p < 0.001 | 0.056 (Negligible) |
| Missing | 0 (0%) | 0 (0%) | 0 (0%) | | | | | |
| **Influence of light exercise over tinnitus** | | | | Chi2 = 320.8** | p < 0.001 | 0.115 (Small) | | |
| Worsens | 231 (16.5%) | 284 (6.8%) | 59 (11.6%) | | | TW VS NE | p < 0.001 | 0.113 (Small) |
| No effect | 937 (66.7%) | 3424 (81.4%) | 278 (54.8%) | | | TI VS NE | p < 0.001 | 0.13 (Small) |
| Improves | 236 (16.8%) | 496 (11.8%) | 170 (33.5%) | | | TW VS TI | p < 0.001 | 0.072 (Small) |
| Missing | 0 (0%) | 0 (0%) | 0 (0%) | | | | | |
| **Somatosensory change : pressing the jaw on the side** | | | | Chi2 = 48.3** | p < 0.001 | 0.063 (Negligible) | | |
| No | 1121 (79.8%) | 3670 (87.3%) | 422 (83.2%) | | | TW VS NE | p < 0.001 | 0.087 (Negligible) |
| Yes | 283 (20.2%) | 534 (12.7%) | 85 (16.8%) | | | TI VS NE | 0.564 | 0.032 (Negligible) |
| Missing | 0 (0%) | 0 (0%) | 0 (0%) | | | TW VS TI | 1.0 | 0.02 (Negligible) |
| **Somatosensory change : pushing jaw backwards** | | | | Chi2 = 24.2** | p < 0.001 | 0.044 (Negligible) | | |
| No | 1151 (82%) | 3667 (87.2%) | 439 (86.6%) | | | TW VS NE | p < 0.001 | 0.062 (Negligible) |
| Yes | 253 (18%) | 537 (12.8%) | 68 (13.4%) | | | TI VS NE | 1.0 | 0.004 (Negligible) |
| Missing | 0 (0%) | 0 (0%) | 0 (0%) | | | TW VS TI | 0.837 | 0.03 (Negligible) |

| | | | | | | | | | |
|---|---|---|---|---|---|---|---|---|---|
| **Somatosensory change : Pushing the jaw outwards rapidly** | | | | Chi2 = 37.8** | p < 0.001 | 0.056 (Negligible) | | | |
| No | 1022 (72.8%) | 3385 (80.5%) | 404 (79.7%) | | | | TW VS NE | p < 0.001 | 0.078 (Negligible) |
| Yes | 382 (27.2%) | 819 (19.5%) | 103 (20.3%) | | | | TI VS NE | 1.0 | 0.005 (Negligible) |
| Missing | 0 (0%) | 0 (0%) | 0 (0%) | | | | TW VS TI | 0.131 | 0.038 (Negligible) |
| **Somatosensory change : Pushing your hand against your forehead while resisting with the neck muscles** | | | | Chi2 = 22.9** | p < 0.001 | 0.043 (Negligible) | | | |
| No | 1094 (77.9%) | 3513 (83.6%) | 418 (82.4%) | | | | TW VS NE | p < 0.001 | 0.061 (Negligible) |
| Yes | 310 (22.1%) | 691 (16.4%) | 89 (17.6%) | | | | TI VS NE | 1.0 | 0.007 (Negligible) |
| Missing | 0 (0%) | 0 (0%) | 0 (0%) | | | | TW VS TI | 1.0 | 0.027 (Negligible) |
| **Somatosensory change : clenching teeth** | | | | Chi2 = 32.1** | p < 0.001 | 0.051 (Negligible) | | | |
| No | 1006 (71.7%) | 3313 (78.8%) | 377 (74.4%) | | | | TW VS NE | p < 0.001 | 0.07 (Negligible) |
| Yes | 398 (28.3%) | 891 (21.2%) | 130 (25.6%) | | | | TI VS NE | 0.932 | 0.029 (Negligible) |
| Missing | 0 (0%) | 0 (0%) | 0 (0%) | | | | TW VS TI | 1.0 | 0.014 (Negligible) |
| **Somatosensory change : Tilting your head backwards** | | | | Chi2 = 34.7** | p < 0.001 | 0.053 (Negligible) | | | |
| No | 1132 (80.6%) | 3656 (87%) | 424 (83.6%) | | | | TW VS NE | p < 0.001 | 0.074 (Negligible) |
| Yes | 272 (19.4%) | 548 (13%) | 83 (16.4%) | | | | TI VS NE | 1.0 | 0.026 (Negligible) |
| Missing | 0 (0%) | 0 (0%) | 0 (0%) | | | | TW VS TI | 1.0 | 0.018 (Negligible) |
| **Somatosensory change : no change with any of these actions** | | | | Chi2 = 62.4** | p < 0.001 | 0.071 (Small) | | | |
| No | 734 (52.3%) | 1703 (40.5%) | 241 (47.5%) | | | | TW VS NE | p < 0.001 | 0.098 (Negligible) |
| Yes | 670 (47.7%) | 2501 (59.5%) | 266 (52.5%) | | | | TI VS NE | 0.132 | 0.038 (Negligible) |
| Missing | 0 (0%) | 0 (0%) | 0 (0%) | | | | TW VS TI | 1.0 | 0.023 (Negligible) |
| **Jaw : my jaw sometimes feels painful** | | | | Chi2 = 1.7 | 1.0 | 0.012 (Negligible) | | | |
| No | 1238 (88.2%) | 3759 (89.4%) | 450 (88.8%) | | | | | | |
| Yes | 166 (11.8%) | 445 (10.6%) | 57 (11.2%) | | | | | | |
| Missing | 0 (0%) | 0 (0%) | 0 (0%) | | | | | | |
| **Jaw : I struggle to fully move my jaw** | | | | Chi2 = 4.3 | 1.0 | 0.019 (Negligible) | | | |
| No | 1358 (96.7%) | 4107 (97.7%) | 496 (97.8%) | | | | | | |
| Yes | 46 (3.3%) | 97 (2.3%) | 11 (2.2%) | | | | | | |
| Missing | 0 (0%) | 0 (0%) | 0 (0%) | | | | | | |
| **Jaw : I struggle to fully move my jaw** | | | | Chi2 = 13.4* | 0.064 | 0.033 (Negligible) | | | |
| No | 1277 (91%) | 3877 (92.2%) | 444 (87.6%) | | | | | | |
| Yes | 127 (9%) | 327 (7.8%) | 63 (12.4%) | | | | | | |
| Missing | 0 (0%) | 0 (0%) | 0 (0%) | | | | | | |
| **Jaw : muscles around my jaw feel tight or tense** | | | | Chi2 = 6.3* | 1.0 | 0.023 (Negligible) | | | |
| No | 1171 (83.4%) | 3612 (85.9%) | 424 (83.6%) | | | | | | |
| Yes | 233 (16.6%) | 592 (14.1%) | 83 (16.4%) | | | | | | |
| Missing | 0 (0%) | 0 (0%) | 0 (0%) | | | | | | |
| **Jaw : I have several popping and clicking noises in my jaw** | | | | Chi2 = 12.7* | 0.087 | 0.032 (Negligible) | | | |
| No | 1104 (78.6%) | 3479 (82.8%) | 406 (80.1%) | | | | | | |
| Yes | 300 (21.4%) | 725 (17.2%) | 101 (19.9%) | | | | | | |

| | | | | | | |
|---|---|---|---|---|---|---|
| Missing | 0 (0%) | 0 (0%) | 0 (0%) | | | |
| **Jaw : I have been diagnosed with TMJ dysfunction** | | | | Chi2 = 3.2 | 1.0 | 0.016 (Negligible) |
| No | 1296 (92.3%) | 3925 (93.4%) | 465 (91.7%) | | | |
| Yes | 108 (7.7%) | 279 (6.6%) | 42 (8.3%) | | | |
| Missing | 0 (0%) | 0 (0%) | 0 (0%) | | | |
| **Jaw : No real issues that I am aware of** | | | | Chi2 = 9.6* | 0.358 | 0.028 (Negligible) |
| No | 530 (37.7%) | 1415 (33.7%) | 191 (37.7%) | | | |
| Yes | 874 (62.3%) | 2789 (66.3%) | 316 (62.3%) | | | |
| Missing | 0 (0%) | 0 (0%) | 0 (0%) | | | |
| **Bruxism : I grind my teeth during my sleep** | | | | Chi2 = 1.0 | 1.0 | 0.009 (Negligible) |
| No | 960 (68.4%) | 2898 (68.9%) | 339 (66.9%) | | | |
| Yes | 444 (31.6%) | 1306 (31.1%) | 168 (33.1%) | | | |
| Missing | 0 (0%) | 0 (0%) | 0 (0%) | | | |
| **Bruxism : I often clench my teeth without realizing it** | | | | Chi2 = 7.8* | 0.751 | 0.025 (Negligible) |
| No | 1027 (73.1%) | 3105 (73.9%) | 345 (68%) | | | |
| Yes | 377 (26.9%) | 1099 (26.1%) | 162 (32%) | | | |
| Missing | 0 (0%) | 0 (0%) | 0 (0%) | | | |
| **Neck stiffness : after certain physical activity** | | | | Chi2 = 2.9 | 1.0 | 0.015 (Negligible) |
| No | 1177 (83.8%) | 3569 (84.9%) | 417 (82.2%) | | | |
| Yes | 227 (16.2%) | 635 (15.1%) | 90 (17.8%) | | | |
| Missing | 0 (0%) | 0 (0%) | 0 (0%) | | | |
| **Neck stiffness : from bad posture** | | | | Chi2 = 11.2* | 0.175 | 0.03 (Negligible) |
| No | 1062 (75.6%) | 3332 (79.3%) | 380 (75%) | | | |
| Yes | 342 (24.4%) | 872 (20.7%) | 127 (25%) | | | |
| Missing | 0 (0%) | 0 (0%) | 0 (0%) | | | |
| **Neck stiffness : I have an associated medical condition** | | | | Chi2 = 0.2 | 1.0 | 0.004 (Negligible) |
| No | 1259 (89.7%) | 3753 (89.3%) | 454 (89.5%) | | | |
| Yes | 145 (10.3%) | 451 (10.7%) | 53 (10.5%) | | | |
| Missing | 0 (0%) | 0 (0%) | 0 (0%) | | | |
| **Neck stiffness : from lying in bed / sleeping** | | | | Chi2 = 7.4* | 0.876 | 0.025 (Negligible) |
| No | 1077 (76.7%) | 3364 (80%) | 395 (77.9%) | | | |
| Yes | 327 (23.3%) | 840 (20%) | 112 (22.1%) | | | |
| Missing | 0 (0%) | 0 (0%) | 0 (0%) | | | |
| **Neck stiffness : my neck movement is restricted due to stiffness** | | | | Chi2 = 4.1 | 1.0 | 0.018 (Negligible) |
| No | 1182 (84.2%) | 3523 (83.8%) | 408 (80.5%) | | | |
| Yes | 222 (15.8%) | 681 (16.2%) | 99 (19.5%) | | | |
| Missing | 0 (0%) | 0 (0%) | 0 (0%) | | | |
| **Neck stiffness : No more than I believe is normal** | | | | Chi2 = 10.1* | 0.293 | 0.029 (Negligible) |
| No | 797 (56.8%) | 2190 (52.1%) | 281 (55.4%) | | | |
| Yes | 607 (43.2%) | 2014 (47.9%) | 226 (44.6%) | | | |
| Missing | 0 (0%) | 0 (0%) | 0 (0%) | | | |

| | | | | | | | | | |
|---|---|---|---|---|---|---|---|---|---|
| **Headaches** | | | | Chi2 = 12.4* | 0.101 | 0.032 (Negligible) | | | |
| No | 821 (58.5%) | 2667 (63.4%) | 302 (59.6%) | | | | | | |
| Yes | 583 (41.5%) | 1537 (36.6%) | 205 (40.4%) | | | | | | |
| Missing | 0 (0%) | 0 (0%) | 0 (0%) | | | | | | |
| **Headaches coming from the neck** | | | | Chi2 = 2.5 | 1.0 | 0.014 (Negligible) | | | |
| No | 1100 (78.3%) | 3369 (80.1%) | 398 (78.5%) | | | | | | |
| Yes | 304 (21.7%) | 835 (19.9%) | 109 (21.5%) | | | | | | |
| Missing | 0 (0%) | 0 (0%) | 0 (0%) | | | | | | |
| **Headaches coming from the jaw** | | | | Chi2 = 8.8* | 0.497 | 0.027 (Negligible) | | | |
| No | 1281 (91.2%) | 3908 (93%) | 456 (89.9%) | | | | | | |
| Yes | 123 (8.8%) | 296 (7%) | 51 (10.1%) | | | | | | |
| Missing | 0 (0%) | 0 (0%) | 0 (0%) | | | | | | |
| **Ear fullness : after activity mainly** | | | | Chi2 = 28.4** | p < 0.001 | 0.048 (Negligible) | | | |
| No | 1309 (93.2%) | 4058 (96.5%) | 480 (94.7%) | | | | TW VS NE | p < 0.001 | 0.066 (Negligible) |
| Yes | 95 (6.8%) | 146 (3.5%) | 27 (5.3%) | | | | TI VS NE | 1.0 | 0.025 (Negligible) |
| Missing | 0 (0%) | 0 (0%) | 0 (0%) | | | | TW VS TI | 1.0 | 0.013 (Negligible) |
| **Ear fullness : after a bad sleep** | | | | Chi2 = 72.7** | p < 0.001 | 0.077 (Small) | | | |
| No | 1250 (89%) | 3994 (95%) | 452 (89.2%) | | | | TW VS NE | p < 0.001 | 0.1 (Small) |
| Yes | 154 (11%) | 210 (5%) | 55 (10.8%) | | | | TI VS NE | p < 0.001 | 0.068 (Negligible) |
| Missing | 0 (0%) | 0 (0%) | 0 (0%) | | | | TW VS TI | 1.0 | 0.0 (Negligible) |
| **Ear fullness : after listening to some sounds or being exposed to noise** | | | | Chi2 = 12.4* | 0.1 | 0.032 (Negligible) | | | |
| No | 1214 (86.5%) | 3760 (89.4%) | 436 (86%) | | | | | | |
| Yes | 190 (13.5%) | 444 (10.6%) | 71 (14%) | | | | | | |
| Missing | 0 (0%) | 0 (0%) | 0 (0%) | | | | | | |
| **Ear fullness : after working at a computer or desk** | | | | Chi2 = 22.5** | p < 0.001 | 0.043 (Negligible) | | | |
| No | 1355 (96.5%) | 4101 (97.5%) | 476 (93.9%) | | | | TW VS NE | 1.0 | 0.025 (Negligible) |
| Yes | 49 (3.5%) | 103 (2.5%) | 31 (6.1%) | | | | TI VS NE | p < 0.001 | 0.058 (Negligible) |
| Missing | 0 (0%) | 0 (0%) | 0 (0%) | | | | TW VS TI | 0.706 | 0.031 (Negligible) |
| **Ear fullness : after periods of stress / anxiety** | | | | Chi2 = 85.3** | p < 0.001 | 0.084 (Small) | | | |
| No | 1129 (80.4%) | 3700 (88%) | 386 (76.1%) | | | | TW VS NE | p < 0.001 | 0.091 (Negligible) |
| Yes | 275 (19.6%) | 504 (12%) | 121 (23.9%) | | | | TI VS NE | p < 0.001 | 0.094 (Negligible) |
| Missing | 0 (0%) | 0 (0%) | 0 (0%) | | | | TW VS TI | 1.0 | 0.025 (Negligible) |
| **Ear fullness : yes but no identified cause** | | | | Chi2 = 2.6 | 1.0 | 0.015 (Negligible) | | | |
| No | 882 (62.8%) | 2549 (60.6%) | 318 (62.7%) | | | | | | |
| Yes | 522 (37.2%) | 1655 (39.4%) | 189 (37.3%) | | | | | | |
| Missing | 0 (0%) | 0 (0%) | 0 (0%) | | | | | | |
| **No ear fullness** | | | | Chi2 = 42.3** | p < 0.001 | 0.059 (Negligible) | | | |
| No | 907 (64.6%) | 2386 (56.8%) | 343 (67.7%) | | | | TW VS NE | p < 0.001 | 0.066 (Negligible) |
| Yes | 497 (35.4%) | 1818 (43.2%) | 164 (32.3%) | | | | TI VS NE | p < 0.001 | 0.059 (Negligible) |
| Missing | 0 (0%) | 0 (0%) | 0 (0%) | | | | TW VS TI | 1.0 | 0.015 (Negligible) |